\documentclass[12pt]{article}
\usepackage{amsfonts}
\usepackage{amsmath}
\usepackage{amssymb}
\usepackage{amsthm}
\newtheorem{assertion}{Conclusion}
\usepackage[utf8]{inputenc}
\usepackage[margin=2cm]{geometry}
\usepackage{geometry}
\usepackage{hyperref}
\usepackage{graphicx}
\usepackage{amsmath}
\usepackage{cite}
\usepackage{hyperref}
\usepackage{tensor}
\usepackage{cases}
\usepackage{appendix}
\usepackage{multirow}
\usepackage{booktabs}
\usepackage{color}
\usepackage[font=small]{subfig}
\usepackage[skip=8pt,labelfont={bf,sc},textfont=small]{caption}

\def\[#1\]{\begin{align}#1\end{align}}
\def \nn {\nonumber}
\def \a {\alpha}
\def \b {\beta}
\def \g {\gamma}

\def \r {\rho}

\def \dd{\mathrm{d}}

\def \e {\epsilon}

\def \f {\phi}
\def \m {\mu}
\def \n {\nu}

\def \pd{\partial}

\def \l{\lambda}

\def \t{\tau}

\def \k{\kappa}
\def \lct{\mathcal{L}_{\text{c.t.}}}
\def \rreg{r_{\text{reg.}}}
\def \rh{r_h}
\def \phih{\phi_h}
\def \phireg{\phi_{\text{reg.}}}
\def \ekr{\text{e}^{\k r^*(r)}}

\def \mo{\mathcal{O}}
\def \e{\text{e}}
\def \hu{\hat{u}}
\def \hv{\hat{v}}
\def \ekr{\text{e}^{\k r^*(r)}}
\def \limrh{\lim\limits_{r\rightarrow r_{h}}}
\def \GD{G_{(D)}}
\def \YMAXFLAT{\max\limits_{\rh<r<r_b}[Y(r)]}

\begin{document}
\begin{titlepage}
\vspace{0.5cm}
\begin{center}
{\Large \bf {The universality of islands outside the horizon}}
\lineskip .75em
\vskip 2.5cm
{\large  Song He$^{a,b,}$\footnote{hesong@jlu.edu.cn}, Yuan Sun$^{a,}$\footnote{sunyuan@jlu.edu.cn} , Long Zhao$^{a,c,d,}$\footnote{zhaolong@mail.itp.ac.cn}, Yu-Xuan Zhang$^{a,}$\footnote{yuxuanz18@mails.jlu.edu.cn}}
\vskip 2.5em
{\normalsize\it $^{a}$Center for Theoretical Physics and College of Physics, Jilin University,\\ Changchun 130012, People's Republic of China\\
 $^{b}$Max Planck Institute for Gravitational Physics (Albert Einstein Institute),\\
Am M\"uhlenberg 1, 14476 Golm, Germany\\
$^{c}$CAS Key Laboratory of Theoretical Physics, Institute of Theoretical Physics,
Chinese Academy of Sciences, P.O. Box 2735, Beijing 100190, China\\
$^{d}$School of Physics, University of Chinese Academy of Sciences, Beijing 100049, China
}
\vskip 3.0em
\end{center}
\begin{abstract}

We systematically calculate the quantum extremal surface (QES) associated with Hawking radiation for general $D$-dimensional ($D\geq2$) asymptotically flat (or AdS) eternal black holes using the island formula. We collect the Hawking radiation particles by a non-gravitational bath and find that a QES exists in the near-horizon region outside the black hole when $c\cdot G_{(D)}$ is smaller enough where $c$ is the central charge of the conformal matter and $G_{(D)}$ the $D$-dimensional Newton constant. The locations of the QES in these backgrounds are obtained and the late-time radiation entropy saturates the two times of black hole entropy. Finally, we numerically check that the no island configuration exists once $c\cdot G_{(D)}$ exceeds a certain upper bound in two-dimensional generalized dilaton theories (GDT). When $c\cdot G_{(D)}$ is close to the upper bound, the backreaction of the matter field on the background can not be neglected. We also consider the conditions of existence of the island configuration with the backreaction and prove that the upper bound also exists for the Witten black hole and Weyl-related Witten black hole.

\end{abstract}
\end{titlepage}

\baselineskip=0.7cm

\tableofcontents
\section{Introduction}
The black hole information paradox \cite{Hawking:1976ra,Mathur:2009hf} is one of the most fundamental problems in contemporary physics. Resolving it has been regarded as the crux of understanding quantum gravity.  {According to Hawking's original calculations, the radiation of a black hole behaving like thermal radiation implies that the entanglement entropy outside the black hole is monotonically increasing. This result contradicts the expectation of the unitarity  of the black hole evaporation process, which is commonly reckoned to be  compatible only with the evolution of radiation entropy satisfying the so-called "Page curve" \cite{Page:1993df,Page:1993wv}.} {Whereas, the original calculations  of the entanglement entropy in \cite{Page:1993df,Page:1993wv} depends on a postulate that the Hilbert space is factorizable. Recent research indicates that the bulk locality is absent in the gravitational system and the boundary system encodes all the bulk information \cite{Raju:2021lwh,Chowdhury:2021nxw,Geng:2021hlu,Raju:2020smc,Laddha:2020kvp}. Without the bulk locality, the whole system can not be divided into the black hole and Hawking radiation intrinsically. In this scenario, one can only collect the bulk information at the asymptotic boundary and then get a constant fine-grained entropy \cite{Raju:2020smc,Laddha:2020kvp,Chowdhury:2021nxw}. {However, the bulk locality can be restored by gluing a non-gravitational system, which is called "bath" conventionally, to the black hole\cite{Chowdhury:2021nxw} with transparent boundary conditions, and one can thus calculate the fine-grained entropy of the Hawking radiation absorbed by the bath.
The price of doing so is that the conservation of stress tensor is broken and the graviton obtains mass \cite{Karch:2000ct,Porrati:2001gx,Porrati:2001db,Geng:2020qvw}.}} {Thanks to the break of the stress tensor conservation, in the AdS/CFT literature, {a new approach, known as the island formula, has been applied to compute the radiation entropy of evaporating black holes and yield the  Page curve \cite{Almheiri:2019qdq,Penington:2019npb,Almheiri:2019hni,Almheiri:2020cfm,Almheiri:2019psf}.}}

{The island formula somehow stems from the investigations of the quantum corrections \cite{Barrella:2013wja,Faulkner:2013ana,Engelhardt:2014gca} of the Ryu-Takayanagi (RT) formula  \cite{Ryu:2006bv,Hubeny:2007xt}. It is well known that the RT formula, as a significant crystallization of the AdS/CFT correspondence \cite{Maldacena:1997re,Gubser:1998bc,Witten:1998qj}, provides a powerful holographic way to evaluate the entanglement entropy of boundary conformal field theory (CFT). Nevertheless, the RT formula is a classical formula, as was proposed in \cite{Engelhardt:2014gca}, when one wishes to count the bulk quantum effects, it should give way to the QES prescription. The QES extremizes the generalized entropy which is the sum of area and bulk entanglement entropy. In terms of the prescription of minimal quantum extremal surface, the island formula for computing the fine-grained entanglement entropy of the Hawking radiation is proposed as\cite{Almheiri:2019hni}}
\[
S_{\text{Rad}}(A)=\min\left\{\mathop{\text{ext}}\limits_I\bigg[\frac{\text{Area}(\pd I)}{4G_N}+S_{\text{matter}}(A\cup I)\bigg]\right\}.\label{Island formula}
\]
{Here $S_{\text{Rad}}(A)$ is the generalized entropy for the radiation in the region $A$, $I$ called island is a bulk region whose boundary $\pd I$ is the minimal quantum extremal surface.} The entanglement entropy from matter part contains the UV divergence which is proportional to the island area, subject to a UV cut-off scale \cite{Bombelli:1986rw,Srednicki:1993im}, and the Newton constant $G_{N}$ must be renormalized \cite{Susskind:1994sm}. The $S_{\text{matter}}$ corresponds to the finite contribution of the matter entanglement entropy. {The validity of this formula is provided by the bulk locality,  thus the coupling of the non-gravitational bath is necessary\cite{Geng:2021hlu}}. Note that \eqref{Island formula} can be also derived from the replica trick for gravitational theories \cite{Almheiri:2019qdq,Penington:2019kki}.


Although the island formula was originally used  to reproduce the Page curve of the evaporating black hole in Jackiw-Teitelboim (JT) gravity \cite{Almheiri:2019hni,Almheiri:2019psf}, the correlational research has been extended to many aspects so far. As an incomplete summary, except for the well-known doubly holographic model as well as the replica wormhole \cite{Krishnan:2020fer,Geng:2020qvw,Caceres:2020jcn,Geng:2021wcq,Geng:2021iyq}, for instance, more on evaporation models and details are explored in \cite{Gautason:2020tmk,Hartman:2020swn,Hollowood:2020cou,Goto:2020wnk,Chen:2020jvn,Wang:2021mqq}. Meanwhile, higher dimensional black hole cases are considered in  \cite{Almheiri:2019psy,Hashimoto:2020cas,Wang:2021woy,Yu:2021cgi,Ahn:2021chg,Karananas:2020fwx,Lu:2021gmv,Krishnan:2020oun} as well as higher derivative gravity \cite{Alishahiha:2020qza,Hernandez:2020nem}. Interestingly, the page curve can be realized in the moving mirror scenario \cite{Akal:2020twv,Reyes:2021npy,Akal:2021foz}, and other quantum information or thermodynamic quantities except entanglement entropy are investigated within island formula   \cite{Kawabata:2021hac,KumarBasak:2020ams,Choudhury:2020hil,Ling:2021vxe,Li:2021dmf,Chandrasekaran:2020qtn,Kawabata:2021vyo,Saha:2021ohr,Pedraza:2021cvx}.  

As pointed in \cite{Almheiri:2019yqk}, within the framework of the so-called "black hole couples thermal baths" model,  the island appears outside the horizon for an external black hole in 2D JT gravity. The radiation entropy approaches $2S_{\text{BH}}$ in the late time limit. There were several case-by-case studies, to confirm the above behavior of QES with the approximation that the central charge of thermal bath is smaller than the inverse of Newton constant associated with a black hole. In this paper, we would like to systematically study QES for various two-dimensional external black holes including asymptotically flat and AdS cases, and higher-dimensional cases. {In asymptotically AdS cases, we couple a flat bath at the boundary of the spacetime, while in asymptotically flat cases, we couple the flat bath at some finite location and then cut off the spacetime region outside it. The Page curves in these kind of models display the information transformation from the gravitational system to the flat bath carried by the Hawking radiation. As mentioned above, coupling a flat bath makes the graviton massive. Unfortunately, the validity of the QES formula and the entanglement wedge reconstruction in both asymptotically flat spacetime and massive gravity theory is still an open question. The island formula and Page curve have been investigated in asymptotically flat spacetime without non-gravitational bath\cite{Hartman:2020swn,Wang:2021woy,Ahn:2021chg,Yu:2021cgi,Karananas:2020fwx,Lu:2021gmv,Gautason:2020tmk,Krishnan:2020oun,Anegawa:2020ezn,Hashimoto:2020cas,Kim:2021gzd}. In this work, we assume the QES formula is applicable in the asymptotically flat spacetime with massive graviton.} In these generic gravitational backgrounds, we try to extract universal features for the existence of QES and islands. We find that once the combination $c\cdot G_{(D)}$ of central charge and Newton constant stays within a certain region, the QES and island configuration in such generic gravitational background always exists outside nearby the black hole event horizon, not inside the horizon. We further do the analytical and numerical self-consistency checks in several GDT.

The organization of this paper is as follows. In Section~\ref{sec:2}, we set up the generic "black hole couples thermal baths" model and obtain certain constraints in terms of the existence of QES. To close this section, we do the numerically self-consistency checks and go beyond the $c\cdot\GD\ll1$ limit in 2D eternal black holes. The summary and prospect are given in section~\ref{sec:3}. Some calculation details and useful formulae are presented in the appendices.

\section{Island formula in  eternal black holes}\label{sec:2}
\subsection{Setup and assumptions}
Let us  consider a $D$-dimensional ($D\geq2$) gravitational system, which consists of  a non-extremal  asymptotically flat (or AdS)  black hole and a thermal bath with which it reaches thermal equilibrium. The whole system is assumed to  be filled with conformal matter with central charge $c$, and the black hole's metric is assumed under the Schwarzschild gauge as follows
\[
\dd s^2=-f(r)\dd t^2+f(r)^{-1}\dd r^2+r^2\dd\Omega^2_{D-2}\label{Schwarzschild gauge}.
\] Here $\dd \Omega^2_{D-2}$ is the unit metric on $\mathbb{S}^{\text{D}-2}$ and $f(r)$ is allowed to have multiple roots and $r_h$ ($f(\rh)=0$) represents the largest one (i.e., the location of the outermost horizon). The black hole's Hawking temperature and entropy are
\[
T_{\text{H}}=\frac{\k}{2\pi}=\frac{f'(\rh)}{4\pi},\quad S=\frac{A(\rh)}{4G_{(D)}},\label{wald entropy}
\]
respectively. Thereinto, $\k$ is surface gravity of the outermost horizon, $G_{(D)}$ is $D$-dimensional Newton constant, and $A(r)$ is a model-dependent function which stands for the area of the $(D-2)$-sphere at radius $r$ in $D\geq3$ dimensional Einstein gravity and represents the value of the dilaton field at $r$ in two-dimensional dilaton gravity \cite{Grumiller:2002nm}, etc.

\begin{figure}[htbp]
	\centering
\captionsetup[subfloat]{farskip=0.1pt,captionskip=5pt}
\subfloat[t][\centering{Asymptotically flat black hole$+$Flat thermal baths}]{\label{Flat BH}
			\includegraphics[width =0.47\linewidth]{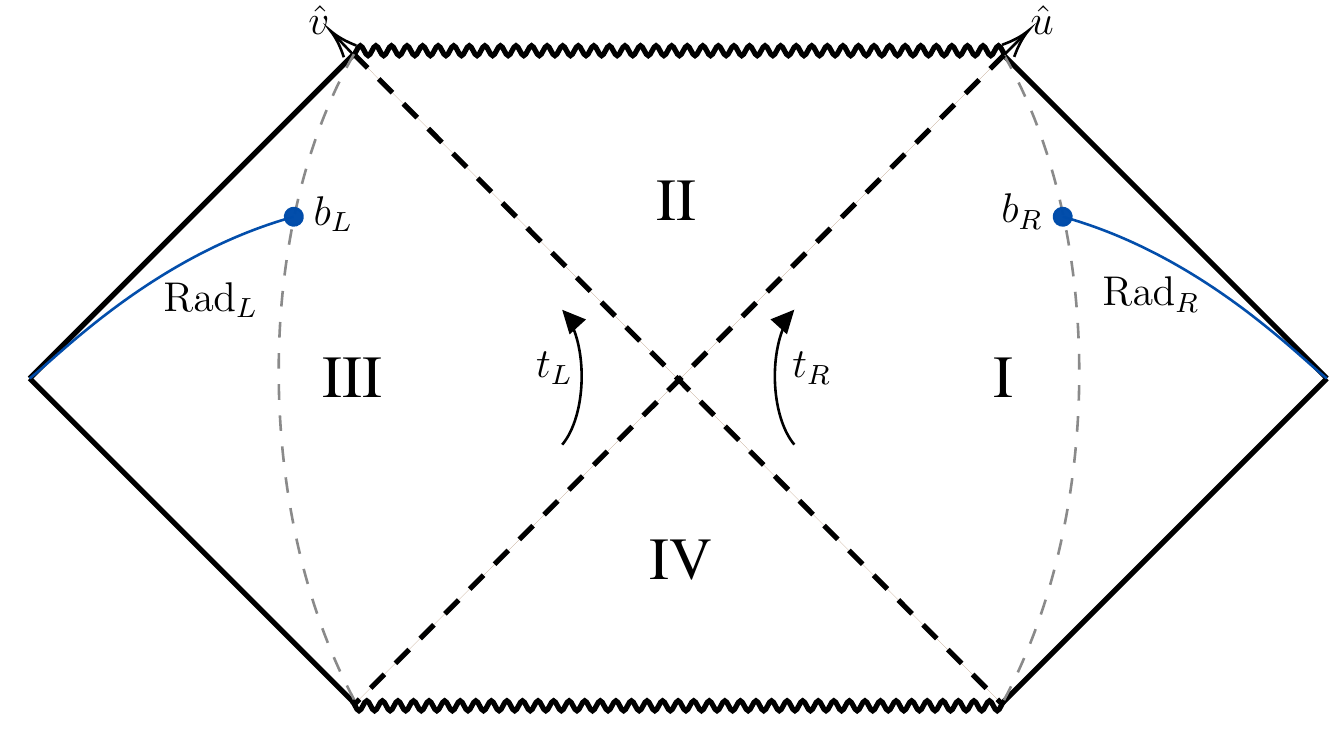}}
\hfill
\subfloat[t][\centering{Asymptotically AdS black hole$+$Flat thermal baths}]{\label{AdS BH}
			\includegraphics[width =0.47\linewidth]{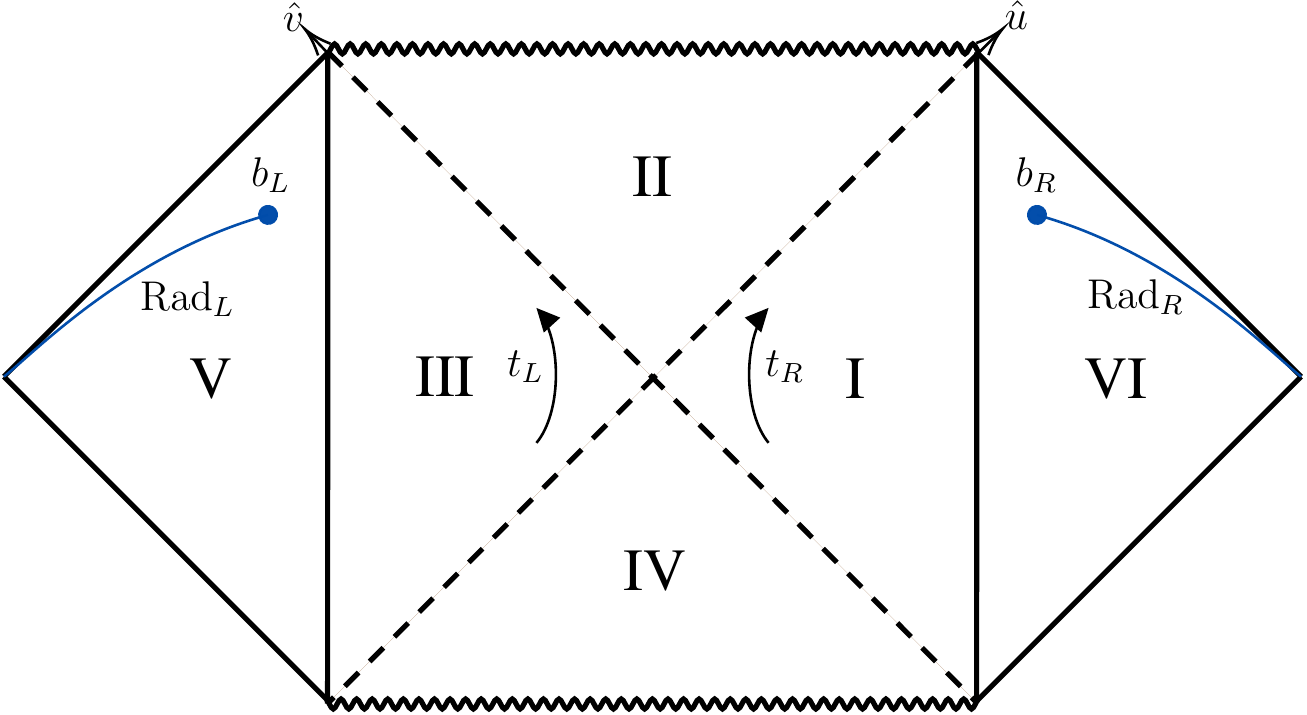}}

	\caption{Penrose diagrams of the whole gravitational system (\textit{Left}: Asymptotically flat black hole with single horizon and singularity. \textit{right}: Asymptotically AdS black hole with single horizon and singularity). Each point on diagrams represents a ($D-2$)-dimensional sphere. The dotted gray lines in (\protect\subref*{Flat BH}) are boundaries of {non-gravitational baths which used to collect} the Hawking radiation. The blue lines stand for collecting region with boundaries $b_{L(R)}$ in a schwarzschild time slice. We consider the symmetric case that $t_{b_L}=t_{b_R}=t_b$ and $r^*_{b_L}=r^*_{b_R}=r^*_b$.}
	\label{black hole penrose diagram}
	\vspace{-0.5em}
\end{figure}

The Penrose diagram of the full system might be depicted as Fig.\ref{black hole penrose diagram}, and the coordinate transformations between Kruskal coordinates and Schwarzschild coordinates in the four wedges of the Penrose diagram of the black hole are set to following
\begin{align}
\text{I}:   &\quad \hat{u}=\k^{-1}\text{e}^{\k(t_R+r^*(r_R))},  & \hat{v}=&-\k^{-1}\text{e}^{-\k(t_R-r^*(r_R))}  & (r_R&>r_h),\\
\text{II}:  &\quad \hat{u}=\k^{-1}\text{e}^{\k(t_R+r^*(r_R))},  & \hat{v}=&~\k^{-1}\text{e}^{-\k(t_R-r^*(r_R))}   & (r_R&<r_h),\\
\text{III}: &\quad \hat{u}=-\k^{-1}\text{e}^{-\k(t_L-r^*(r_L))},& \hat{v}=&~\k^{-1}\text{e}^{\k(t_L+r^*(r_L))}    & (r_L&>r_h),\\
\text{IV}:  &\quad \hat{u}=-\k^{-1}\text{e}^{\k(t_L+r^*(r_L))}, & \hat{v}=&-\k^{-1}\text{e}^{-\k(t_L-r^*(r_L))}  & (r_L&<r_h),
\end{align}
where $r^*\equiv\int^{r}f(\tilde{r})^{-1}\dd\tilde{r}$ is tortoise coordinate. The transformations above give the length element in Kruskal coordinates
\[
\dd s^2=-\e^{2\rho}\dd\hu\dd\hv+r_{R(L)}^2\dd\Omega^2_{D-2}\quad\left(\e^{2\rho}\equiv f(r_{R(L)})\e^{-2\k r^{*}(r_{R(L)})}\right).
\]

As shown in Fig.\subref{AdS BH}, we have adopted the customary approach to deal with the black hole and the thermal bath in the case of asymptotically AdS: $D$-dimensional flat spacetimes $\mathbb{R}^{1,\text{D}-1}$ will be used as auxiliary thermal baths to be glued to both sides of the two-sided black hole\cite{Almheiri:2019yqk}\footnote{We follow the prescription in \cite{Almheiri:2019yqk} but generalize it to higher-dimensions. Firstly the tortoise coordinates are normalized by requiring $\lim\limits_{r_{R(L)}\rightarrow\infty}r_{R(L)}^*=0$, such that the right (left) bath corresponds to $r^*_{R(L)}>0$. The Kruskal coordinates thus can be extended to the baths (V and VI):
\[
\text{V}: &\quad \hat{u}=-\k^{-1}\text{e}^{-\k(t_L-r^*_L)},& \hat{v}=&~\k^{-1}\text{e}^{\k(t_L+r^*_L)}    & (r^*_L&>0),\nn\\
\text{VI}: &\quad \hat{u}=\k^{-1}\text{e}^{\k(t_R+r^*_R)},  & \hat{v}=&-\k^{-1}\text{e}^{-\k(t_R-r^*_R)}  & (r^*_R&>0).\nn
\]
Meanwhile, we assume that the two-sided black hole is truncated at $r_{R}=\Lambda$ and $r_{L}=\Lambda$ respectively, and the metric of the right (left) bath is set to following
\[
\dd s^2=f(\Lambda)\left(-\dd t_{R(L)}^2+(\dd r^*_{R(L)})^2\right)+\left(\sqrt{f(\Lambda)} r^*_{R(L)}+\Lambda\right)^2\dd\Omega_{D-2}^2\nn
\]
to ensure that two metrics (black hole and bath) are continuously connected at the cut-off. Note that this metric is flat.
}.
When the spacetime is asymptotically flat, the prevalent method is to select the region far away from the black hole as the thermal bath\cite{Anegawa:2020ezn,Gautason:2020tmk}. As discussed in the introduction, the bulk locality is absent without a flat bath, and as a result the island formula is inapplicable. To solve this problem, we couple a flat bath to collect Hawking radiation at a certain Schwarzschild coordinate $r\rightarrow r_{b+}$, as shown in Fig.\subref{Flat BH}. Note that the dependence on the character of the bath in equation \eqref{s-wave approximation} is only the location $r_b$ and the Weyl factor $W(r_b)$ of $b$. With the benefit of the continuity of the metric at $r_b$, these two quantities of our models are the same as those models without flat bath \cite{Anegawa:2020ezn,Hashimoto:2020cas,Ahn:2021chg,Yu:2021cgi,Karananas:2020fwx,Lu:2021gmv,Nam:2021bml,Wang:2021woy,Kim:2021gzd}.

 It is also important to emphasize that, when $D\geq3$, the $s$-wave approximation \cite{Hemming:2000as,Hashimoto:2020cas} has been taken  into account in the calculations of $S_{\text{matter}}$ below. The entanglement entropy of matter between two shells $S_1$ and $S_2$ becomes
\[
S_{\text{matter}}(S_1,S_2)=&\frac{c}{6}\log d^2(S_1,S_2)\nn\\
=&\frac{c}{6}\log\left\lvert\big(\hat{u}(S_1)-\hat{u}(S_2)\big)\big(\hat{v}(S_1)-\hat{v}(S_2)\big)\sqrt{\text{W}(S_1)\text{W}(S_2)}\right\lvert\label{s-wave approximation},
\]
when the quantum state of total system is vacuum in  $(\hat{u},\hat{v})$ coordinates. In the above, $W(S_1)$ and $W(S_2)$ are warped factors of the metric at $S_1$ and $S_2$ under the $(\hat{u},\hat{v})$ coordinates, respectively.
\subsection{Without island, the radiation entropy diverges linearly}
In this section, we evaluate the entanglement entropy of the Hawking radiation at late times in the missing island construction. It shows that "information loss" is a common phenomenon for black holes we are considering.

 Without  island, the only contribution of \eqref{Island formula} is coming from  the  collecting regions of the Hawking radiation (see $\text{Rad}_L$ and  $\text{Rad}_R$ in Fig.\ref{black hole penrose diagram}). The collecting region on the right (left) is the region outside the shell $r^*_{R(L)}=r^*_{b_{R(L)}}$ in time slices of $(t_{R(L)},r^*_{R(L)})$ coordinates and we shall choose the symmetric configuration $r^*_{b_L}=r^*_{b_R}=r^*_b$ and $t_{b_L}=t_{b_R}=t_b$ in the following calculations. Assuming that the state of total system is  vacuum  in $(\hat{u},\hat{v})$ coordinates, The formula can be further reduced to the entanglement entropy of the interval $[b_L,b_R]$ by \eqref{s-wave approximation}, that is

\[
S_{\text{Rad}}=&\frac{c}{6}\log\Big\lvert\big(\hat{u}(b_L)-\hat{u}(b_{R})\big)\big(\hat{v}(b_{L})-\hat{v}({b_{R}})\big)\sqrt{W(b_L)W(b_R)}\Big\rvert,
\]
where
\begin{equation}   W(b_{R(L)}) =
 \begin{cases}
  -f(r_b)\e^{-2\k r^*_b}, & \text{for asymptotically flat black holes},\\
   -f(\Lambda)\e^{-2\k r^*_b}, & \text{for asymptotically AdS black holes}.\label{W(b)}
 \end{cases}
\end{equation}
Simple calculation shows that
 \[
    S_{\text{Rad}} =& \left\{\begin{array}{lr}
        \frac{c}{6}\log\bigg(4\k^{-2}f(b)\cosh^2\k t_b\bigg), & \text{for asymptotically flat black holes}\nn\\
        \frac{c}{6}\log\bigg(4\k^{-2}f(\Lambda)\cosh^2\k t_b\bigg), & \text{for asymptotically AdS black holes}
        \end{array}\right\}\\
         \simeq&\frac{c}{3}\k t_b+\text{time independent terms}\label{linear growth}.
  \]
Notice that \eqref{linear growth} holds for all black holes we are considering.  The linear growth of radiation entropy when the island contribution is missing obviously contradicts the Page curve and thus leads to the information paradox for the black hole.
\subsection{Island emerges outside the  horizon and saves the entropy bound}
\begin{figure}[htbp]
	\centering
\captionsetup[subfloat]{farskip=0.1pt,captionskip=5pt}
\subfloat[t][\centering{Asymptotically flat black hole with an island$+$Flat thermal baths}]{\label{Flat BH+Island}
			\includegraphics[width =0.47\linewidth]{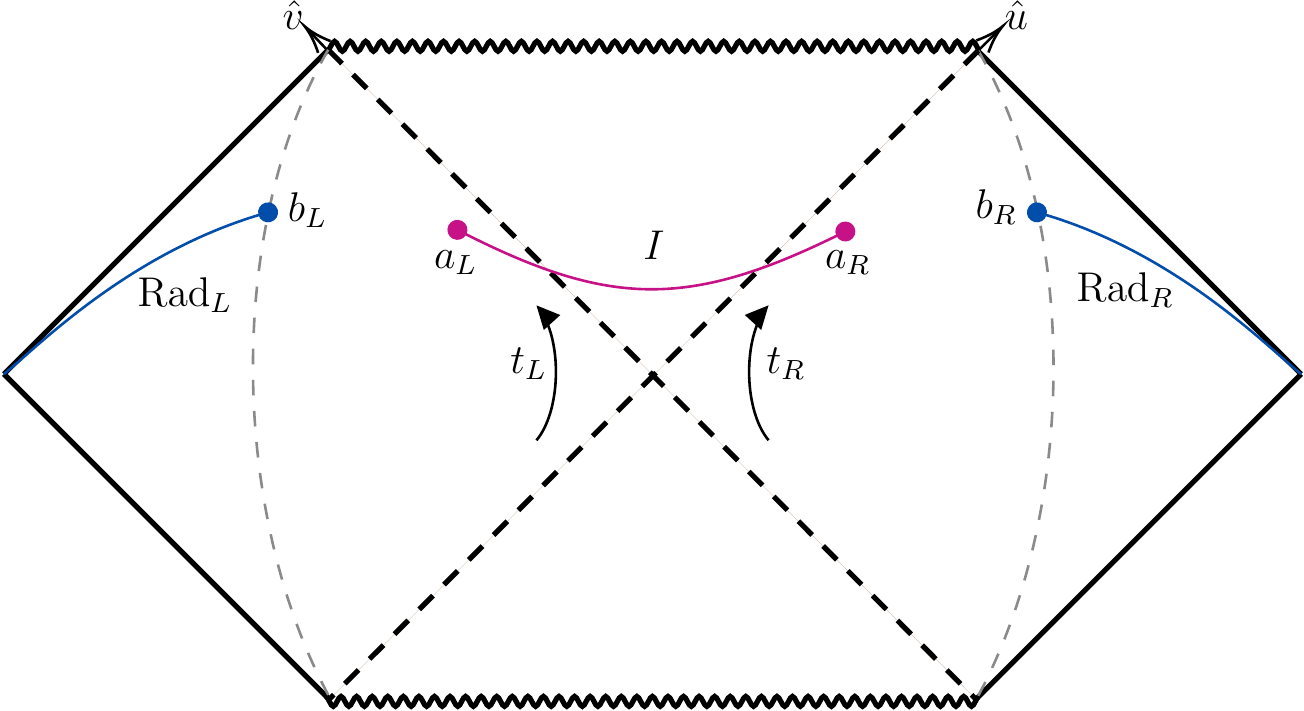}}
\hfill
\subfloat[t][\centering{Asymptotically AdS black hole with an island$+$Flat thermal baths}]{\label{AdS BH+Island}
			\includegraphics[width =0.47\linewidth]{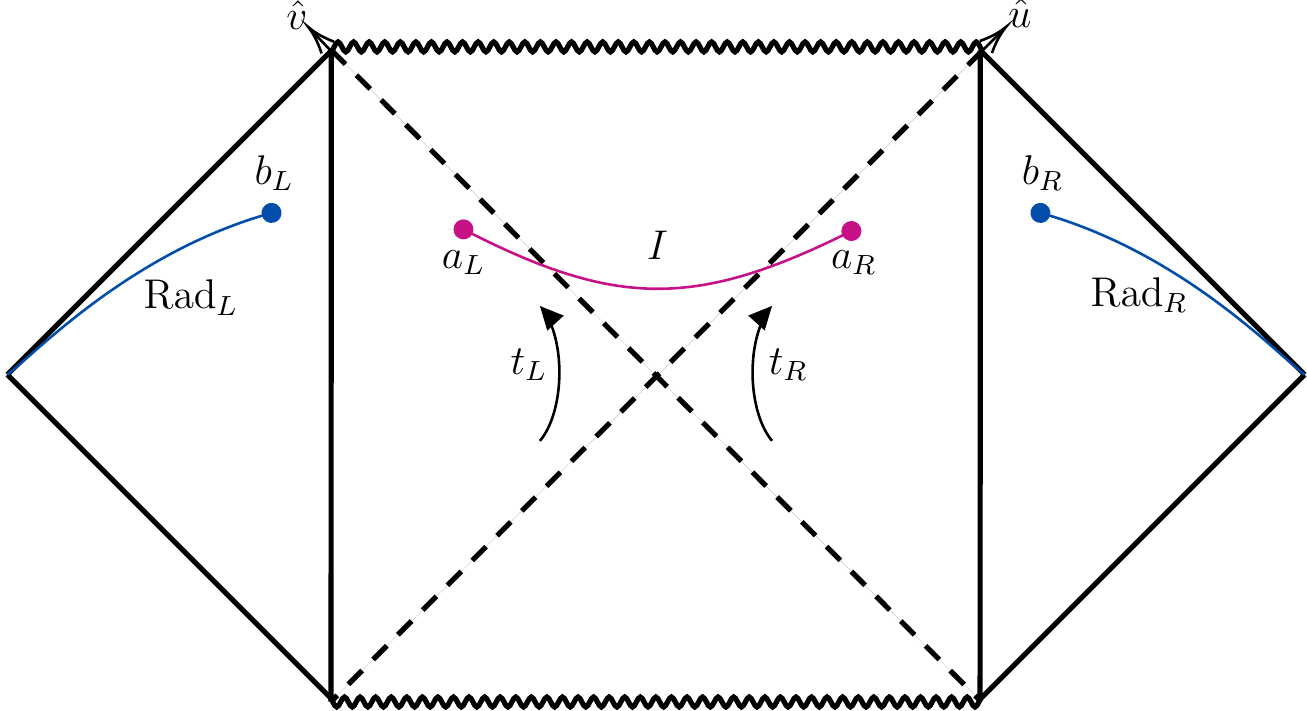}}

	\caption{Penrose diagrams with islands (\textit{Left}: Asymptotically flat black hole with single horizon and singularity. \textit{right}: Asymptotically AdS black hole with single horizon and singularity). The pink lines are islands whose boundaries are outside the horizon. The dotted gray lines in (\protect\subref*{Flat BH+Island}) are boundaries of collecting region for the Hawking radiation. The blue lines stand for collecting region with boundaries $b_{L(R)}$ in a Schwarzschild time slice. We consider the symmetric case that $t_{b_L}=t_{b_R}=t_b$,  $t_{a_L}=t_{a_R}=t_a$, $r_{a_L}=r_{a_R}=r_a$, $r^*_{b_L}=r^*_{b_R}=r^*_b$.}
	\label{diagram with island}
	\vspace{-0.5em}
\end{figure}

In this section, we shall reconsider the entropy of the Hawking radiation by counting the contribution of the island. It is easy to verify that the equation determining the location of QES has no solution inside the horizon. Therefore the basic configuration is set as shown in Fig.\ref{diagram with island}. As shown in Fig.\ref{diagram with island}, we are continuing  with the symmetric structure used in the previous section. The two boundaries of island are marked $a_L$ and $a_R$ respectively, and $t_{a_L}=t_{a_R}=t_a$, $r_{a_L}=r_{a_R}=r_a$.
After taking $s$-wave approximation for $D\geq3$, it shows that the entanglement entropy of conformal matter in $\{\text{Rad}\cup I\}$ can be well approximated by twice of the  entanglement entropy in the single interval $[a_R,b_R]$ when $t_b$ and $t_a\rightarrow\infty$ \cite{Casini:2005rm}
\[
S_{\text{Rad}}=&\frac{A(a_R)}{2G_{(D)}}+\frac{c}{3}\log\bigg\lvert\big(\hat{u}\left({a_R}\right)-\hat{u}\left({b_R}\right)\big)\big(\hat{v}\left({a_R}\right)-\hat{v}\left({b_R}\right)\big)\sqrt{W(a_R)W(b_R)}\bigg\rvert,\label{orginal with-islanded entropy}
\]
where $W(a_R)=-f(r_a)\e^{-2\k r^*_a}$ and $W(b_R)$ is \eqref{W(b)}.
 Eq.\eqref{orginal with-islanded entropy} can be expressed in $(t_R,r_R)$ coordinates
\[
S_{\text{Rad}}=&\frac{c}{3}\log\bigg\lvert\k^{-2}\big(f(r_a)f(r_b)\text{e}^{-2\k(r_a^*+r^*_b)}\big)^{\frac{1}{2}}\bigg(2\text{e}^{\k(r^*_a+r^*_b)}\cosh[\k(t_b-t_a)]-(\text{e}^{2\k r_a^*}+\text{e}^{2\k r^*_b})\bigg)\bigg\lvert+\frac{A(r_a)}{2G_{(D)}},\label{complexed formula for entropy}\\
&\text{for asymptotically flat black holes.}\qquad f(r_b)\rightarrow f(\Lambda)~\text{for asymptotically AdS black holes.}\nn
\]
It's easy to find $t_a$ should be equal to $t_b$ when we extremise $S_{\text{Rad}}$ with respect to $t_a$, then we arrive at  a simpler expression compared to \eqref{complexed formula for entropy},
\[
S_{\text{Rad}}=&\frac{A(r_a)}{2G_{(D)}}+\frac{2c}{3}\log\left[\frac{\text{e}^{\k r^*_b}-\text{e}^{\k r^*_a}}{\k}\right]+\frac{c}{6}\log\left[ f(r_a)f(r_b)\text{e}^{-2\k\big(r^*_a+r^*_b\big)}\right],\label{refined radiation entropy}\\
&\text{for asymptotically flat black holes.}\qquad f(r_b)\rightarrow f(\Lambda)~\text{for asymptotically AdS black holes.}\nn
\]
Taking partial derivative of $S_{\text{Rad}}$ with respect to $r_a$, we meet the algebra equation of determining the location of  QES ($r_a$ here)
\[
\pd_{r_a}S_\text{Rad}=&\frac{A'(r_a)}{2G_{(D)}}-\frac{2c}{3}\frac{\k}{f(r_a)\bigg(\text{e}^{\k(r^*_b-r^*_a)}-1\bigg)}+\frac{c}{6}\frac{f'(r_a)-2\k}{f(r_a)}=0,
\label{QES equation}
\]
which  is the same for both asymptotically flat black holes and asymptotically AdS black holes.
There are  some  model-independent properties of the solution that can be extracted from \eqref{QES equation}, notwithstanding this algebra equation of $r_a$ may be precisely solved only after $f(r)$ and $A(r)$ are given. The key point essentially comes from the fact that the near-horizon geometry is common to all non-extreme  black holes. To show them clearly, let's rewrite \eqref{QES equation} as follows
\[
Y(r)\equiv\frac{3A'(r)}{2}\cdot\left(\frac{2\text{e}^{\k r^*(r)}}{f(r)}\left(\frac{\k}{\text{e}^{\k r^*_b}-\text{e}^{\k r^*(r)}}\right)+\frac{2\k-f'(r)}{2f(r)}\right)^{-1}=c\cdot G_{(D)},\label{cG=Y(a)}
\]
where the subscript $a$ has been omitted for brevity. The zero points of $\pd_{r_a}S_{\text{Rad}}$ now become the points of intersection between the horizontal line $y=c\cdot G_{(D)}$ and the curve $y=Y(r)$ \big($\rh<r<r_b$ for asymptotically flat and $\rh<r<\Lambda$ for asymptotically AdS\big) on the $r-y$ plane, as shown in Fig.\ref{Fig: Y function for flat BH}. Let's  focus on the behavior of $Y(r)$ near $\rh$. A rough estimation can be made since $f(r)\approx 2\k(r-\rh)$ and $r^*(r)\approx \frac{1}{2\k}\log\big[\frac{r}{\rh}-1\big]$ for $r\gtrsim\rh$. $Y(r)$ can thus be approximated to
\[
Y(r)\approx \frac{3}{2}A'(\rh)\Big(X\cdot\rh^{-1}\e^{-\k r^*_b}(\frac{r}{\rh}-1)^{-\frac{1}{2}}-\frac{f''(\rh)}{4\k}\Big)^{-1}\sim\sqrt{\frac{r}{\rh}-1},\label{approximation of Y}
\]
where $X$ is an undetermined constant. The approximate behavior of function $Y$ near $\rh$ is sufficient for us to draw two following conclusions:
\begin{assertion}\label{conclusion 1}
There must be a quantum extremal surface located in the near-horizon region outside the black hole, 
\[
r_a=\rh+    \frac{8\k(c\cdot G_{(D)})^2}{9A'(\rh)^2}\exp\Big\{-2\k r^*_b-2\rho(\rh)\Big\}+\mo\Big(\big(c\cdot G_{(D)}\big)^3\Big),\label{approximate solution}
\]
when $c\cdot G_{(D)}\ll 1$.
\end{assertion}
\begin{assertion}\label{conclusion 2}
There has to be an upper bound on $c\cdot G_{(D)}$ to have an island configuration.
\end{assertion}
     The second conclusion can be a direct corollary to the \textit{boundedness theorem}, since $Y(r)$ is a continuous function on the closed interval $[\rh,r_b]$ ($[\rh,\Lambda]$ for asymptotically AdS).\footnote{It should be emphasized that conclusion \ref{conclusion 2} is based on the fact that we completely ignore the backreaction of Hawking radiation, and we reinvestigate the effect of backreaction  on island configuration in section \ref{section: balckreaction}.} While for the conclusion \ref{conclusion 1}, firstly, the approximate behavior of $Y$ guarantees that when $c\cdot G_{(D)}\ll1$ there must be a point of intersection near $\rh$,  which is graphically obvious.\footnote{One may worry that we may miss some other points of intersection. Indeed, for asymptotically flat black holes, it's not hard to find that there is another intersection near $r_b$, which we call $r_{a'}$. However, when considering the constraint that $c\cdot G_{(D)}\ll1$, the leading order contribution of the island formula comes from the area term, and since $r_{a'}>r_a$, we have $S_{\text{Rad}}(r_{a'})>S_{\text{Rad}}(r_a)$. The root near $r_b$ is thus discarded.} Secondly, the approximate formula \eqref{approximate solution} is obtained by Taylor expansion of the local inverse function of $Y$  near $\rh$.\footnote{For details, please refer to Appendix \ref{appendix 1: derivation of ra}.} As listed in Table \ref{table: list of qes}, we calculate the approximate locations of QESs for several common black holes by \eqref{approximate solution} and compare them with existing results\cite{Anegawa:2020ezn,Hashimoto:2020cas,Yu:2021cgi,Nam:2021bml,Wang:2021woy,Kim:2021gzd}. Note that those results are calculated in the models with gravitational bath which is different from our models. However, the coupling of a non-gravitational bath doesn't affect the result mathematically as discussed in the last subsection.
\begin{table}[]
\centering
\caption{Approximations of location of quantum extremal surface for several black holes}
\label{table: list of qes}
\begin{tabular}{clclclcl}
\toprule[1pt]
\multicolumn{2}{c}{\multirow{2}{*}{Black hole}}      & \multicolumn{2}{c}{\multirow{2}{*}{$A(r)$}}  & \multicolumn{2}{c}{\multirow{2}{*}{$f(r)$}}  & \multicolumn{2}{c}{\multirow{2}{*}{$r_a-r_h\approx$}} \\
\multicolumn{2}{c}{}                         & \multicolumn{2}{c}{}                    & \multicolumn{2}{c}{}                    & \multicolumn{2}{c}{}                       \\\hline
\multicolumn{2}{c}{\multirow{2}{*}{Witten(CGHS)}}  & \multicolumn{2}{c}{\multirow{2}{*}{$\e^{2\l r}$}} & \multicolumn{2}{c}{\multirow{2}{*}{$\mathnormal{\text{1\large{$-\e^{-2\l(r-\rh)}$}}}$}} & \multicolumn{2}{c}{\multirow{2}{*}{$\frac{2c^2G_{(2)}^2}{9\l}\left(\e^{2\l(\rh+r_b)}-\e^{4\l\rh}\right)^{-1}$}}    \\
\multicolumn{2}{c}{}                         & \multicolumn{2}{c}{}                    & \multicolumn{2}{c}{}                    & \multicolumn{2}{c}{}                       \\
\multicolumn{2}{c}{\multirow{2}{*}{JT}}      & \multicolumn{2}{c}{\multirow{2}{*}{$\mathnormal{\text{\large{$\frac{r}{L}$}}}$}} & \multicolumn{2}{c}{\multirow{2}{*}{$\mathnormal{\text{\large{$\frac{r^2-\rh^2}{L^2}$}}}$}} & \multicolumn{2}{c}{\multirow{2}{*}{$\frac{2c^2G_{(2)}^2L^2}{9\rh}\e^{-2\frac{\rh}{L^2}r^*_b}$}}    \\
\multicolumn{2}{c}{}                         & \multicolumn{2}{c}{}                    & \multicolumn{2}{c}{}                    & \multicolumn{2}{c}{}                       \\
\multicolumn{2}{c}{\multirow{2}{*}{BTZ}}     & \multicolumn{2}{c}{\multirow{2}{*}{$2\pi r$}} & \multicolumn{2}{c}{\multirow{2}{*}{$\mathnormal{\text{\large{$\frac{r^2-\rh^2}{L^2}$}}}$}} & \multicolumn{2}{c}{\multirow{2}{*}{$\frac{c^2G_{(3)}^2}{18\pi^2\rh}\e^{-2\frac{\rh}{L^2}r^*_b}$}}    \\
\multicolumn{2}{c}{}                         & \multicolumn{2}{c}{}                    & \multicolumn{2}{c}{}                    & \multicolumn{2}{c}{}                       \\
\multicolumn{2}{c}{\multirow{2}{*}{4d-Schwarzschild}} & \multicolumn{2}{c}{\multirow{2}{*}{$4\pi r^2$}} & \multicolumn{2}{c}{\multirow{2}{*}{$\mathnormal{\text{1\large{$-\frac{\rh}{r}$}}}$}} & \multicolumn{2}{c}{\multirow{2}{*}{$\frac{c^2G_{(4)}^2}{144\pi^2\rh^2(r_b-\rh)}\e^{1-\frac{r_b}{\rh}}$}}    \\
\multicolumn{2}{c}{}                         & \multicolumn{2}{c}{}                    & \multicolumn{2}{c}{}                    & \multicolumn{2}{c}{}                       \\
\multicolumn{2}{c}{\multirow{2}{*}{4d-non-extremal RN}}      & \multicolumn{2}{c}{\multirow{2}{*}{$4\pi r^2$}} & \multicolumn{2}{c}{\multirow{2}{*}{$\mathnormal{\text{(1\large{$-\frac{r_+}{r}$})(\normalsize{$1$}\large{$-\frac{r_-}{r}$})}}$}} & \multicolumn{2}{r}{\multirow{2}{*}{$\frac{c^2G_{(4)}^2}{144\pi^2r_+^2(r_b-r_+)}\left(\frac{r_b-r_-}{r_+-r_-}\right)^{\frac{r_-^2}{r_+^2}}\e^{-\frac{(r_b-r_+)(r_+-r_-)}{r_+^2}}$}}    \\
\multicolumn{2}{c}{}                         & \multicolumn{2}{c}{}                    & \multicolumn{2}{c}{}                    & \multicolumn{2}{c}{}\\ 
\multicolumn{2}{c}{}                         & \multicolumn{2}{c}{}                    & \multicolumn{2}{c}{}                    & \multicolumn{2}{c}{}\\
\bottomrule[1pt]
\end{tabular}
\end{table}
Explicitly, our results coincide with qualitative results in literature \cite{Anegawa:2020ezn}, and exactly match the quantitative results in \cite{Hashimoto:2020cas}.\footnote{In addition, Eq.\eqref{approximate solution} can also reproduce the results in \cite{Yu:2021cgi,Nam:2021bml} and differ by a scale factor from those of \cite{Wang:2021woy,Kim:2021gzd}.}

\begin{figure}[htbp]
  \centering
  \includegraphics[width=0.5\linewidth]{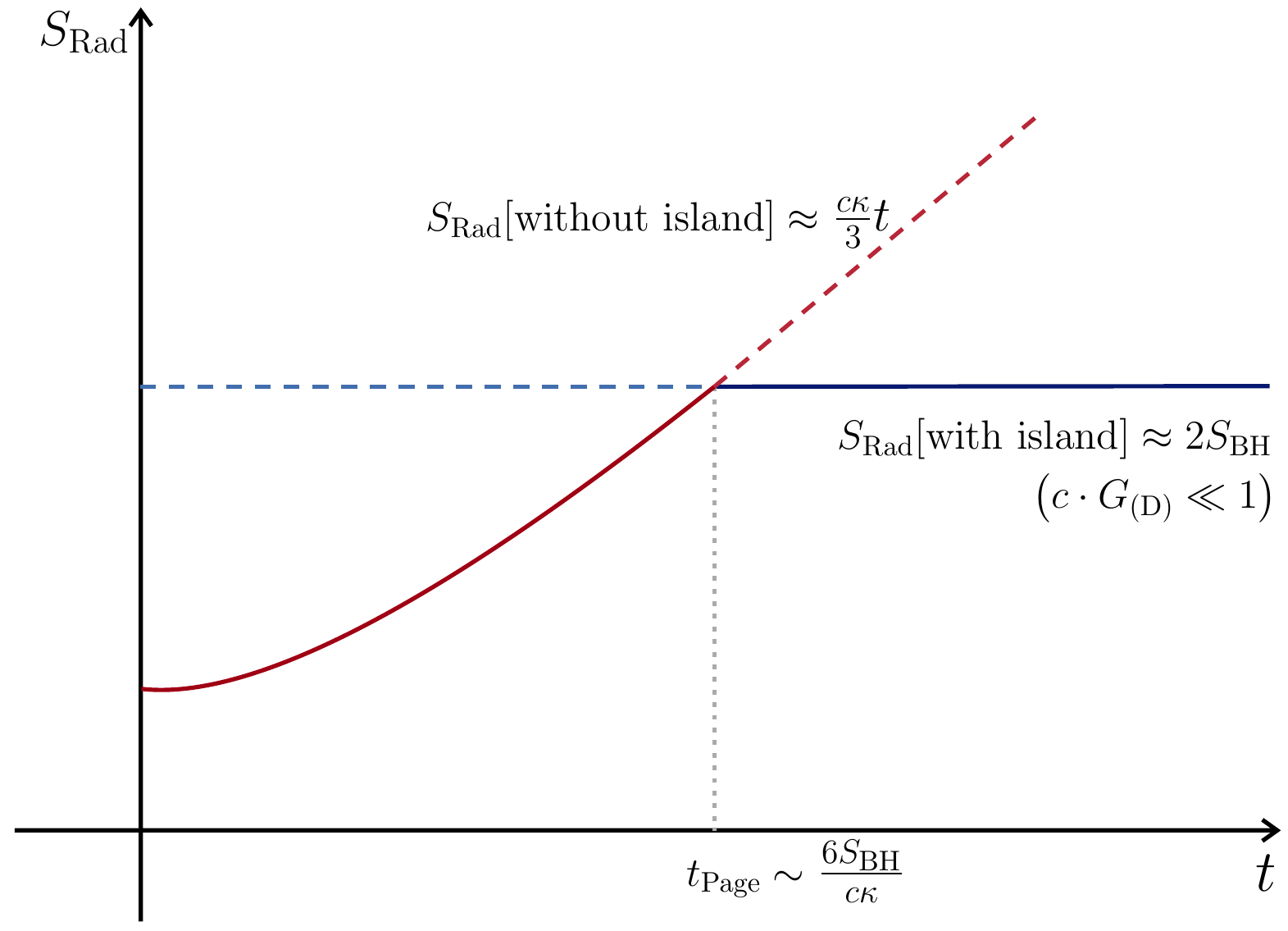}
  \caption{The Page curve for $D$-dimensional non-extremal spherically symmetric black holes. When the contribution of the island is not considered, the late-time radiation entropy increases linearly (red dashed line); After considering the island's contribution and the condition $c\cdot G_{(D)}\ll1$, the late-time radiation entropy is approximatively saturated at $2S_{\text{BH}}$  (blue solid line). }\label{page curve}
\end{figure}

Substituted the approximate solution \eqref{approximate solution} into \eqref{refined radiation entropy}, the late-time radiation entropy after including the island contribution can be obtained as\footnote{Similar to \eqref{approximate solution}, the derivation is a little tricky, please refer to Appendix \ref{appendix2:2SBH} for details.}
\[
S_{\text{Rad}}[\text{with island}]=&\frac{A(\rh)}{2G_{(D)}}+\frac{c}{3}\log d^2(\rh,r_b)-\frac{4\k c^2G_{(D)}}{9A'(\rh)}\exp\big\{-2\k r^*_b-2\rho(r_h)\big\}+\mo(c^3G_{(D)}^2)\nn\\
=&2S_{\text{BH}}+\mo(c)  \quad (c\cdot G_{(D)}\ll1).\label{late-time entropy: with island}
\]
Note that the above approximation formula for the late-time radiation entropy also coincide with  results in \cite{Hashimoto:2020cas,Nam:2021bml}.  Based on above results, we can reproduce the Page curve for generic non-extremal spherically symmetric black holes as Fig.\ref{page curve}. Under the constraint $c\cdot G_{(D)}\ll 1$, the estimation of the Page time also has a concise and uniform form, $t_{\text{Page}}\sim\frac{6S_{\text{BH}}}{c\k}=\frac{3S_{\text{BH}}}{\pi cT_{\text{H}}}$ for all black holes that meet the requirements. Note that, as shown in the next section, once the condition $c\cdot G_{(D)}\ll1$ is broken, the late-time radiation entropy after considering the island contribution does not saturate near $2S_{\text{BH}}$, but has a significant deviation. This suggests that the estimation for the Page time will also change.

\subsection{Go beyond $c\cdot G_{(D)}\ll1$: Examples in two-dimensional dilaton gravity}
In the previous section, we show that when $c\cdot\GD\ll1$, there must be a QES located in the near-horizon region outside the black hole, and the late-time radiation entropy given by it is saturated near $2S_{\text{BH}}$ (with sub-leading corrections of order $c$).  It is natural to ask how does the island change when $c\cdot\GD\ll1$ is no longer satisfied. One can expect that the location of the QES might be model-dependent and the late-time radiation entropy may deviate from $2S_{\text{BH}}$ significantly. In this section, we shall numerically solve the equation \eqref{QES equation} in eternal black hole solutions of two-dimensional GDT to look at the change of the island as $c\cdot G_{(2)}$ varies.
\subsubsection{Eternal black holes in GDT }
The  action of the GDT in 2 dimensions is given by\cite{Grumiller:2002nm}
\[
I_{\text{GDT}}=\frac{1}{16\pi G_{(2)} }\int_{\mathcal{M}}\sqrt{-g}\left(\phi R+U(\phi)\left(\nabla\phi\right)^2+V(\phi)\right)\dd^2x+\frac{1}{8\pi G_{(2)}}\int_{\pd\mathcal{M}}\sqrt{-h} \left(\phi K-\mathcal{L}_{\text{c.t.}}\right)\dd x\label{GDT's action}.
\]
 Note that in the above equation, $U(\phi)$ and $V(\phi)$ are arbitrary functions of dilaton field $\phi$. The boundary term in the action involving the extrinsic curvature $K$ and a counterterm $\lct$\footnote{The exact form of $\lct$ depends on the selection of $U(\phi)$ and $V(\phi)$, please refer to Appendix \ref{thermo} for details. One can also refer to \cite{Cai:2020wpc} for details.} plays two main roles\cite{Grumiller:2007ju}: 1) It makes the variational properties of the action compatible with the semi-classical approximation of the path integral.   2) It renders the Euclidean on-shell action finite and gives the correct black hole thermodynamics.

Given proper functions $U$ and $ V$, one can in principle obtain a series of physically reasonable solutions. Notably, a family of eternal black hole solutions have been given in \cite{Grumiller:2002nm,Grumiller:2007ju}: Under the Schwarzschild gauge of the metric and the time-independent presupposition of the dilaton field
\[
\dd s^2=-f(r)\dd t^2+f(r)^{-1}\dd r^2,\quad \phi=\phi(r),\label{ansatz}
\]
the equations of motion corresponding to \eqref{GDT's action}
\[
\pd_{\phi}U(\phi)\left(\nabla\phi\right)^2+2U(\phi)\nabla^2\phi-\pd_{\phi}V(\phi)=&R,\\
U\nabla_a\phi\nabla_b\phi-\nabla_a\nabla_b\phi+g_{ab}\left[\nabla^2\phi-\frac{1}{2}U(\phi)\left(\nabla\phi\right)^2-\frac{1}{2}V(\phi)\right]=&0
\]
can be solved as
\[
r=&\int^\phi \text{e}^{Q(\phi')}\dd\phi'+C,\label{SolutionPhi}\\
f(r)\equiv F\big(\phi(r)\big)=&\left(W(\phi)-16\pi G_{(2)}M\right)\text{e}^{Q(\phi)},\label{SolutionMetric}
\]
where
\[
Q(\phi)=Q_0-\int^\phi U(\phi')\dd\phi',\label{Q}\\
W(\phi)=W_0+\int^\phi V(\phi')\text{e}^{Q(\phi')}\dd \phi'.\label{W}
\]
Here $C$, $Q_0$, $W_0$ are integration constants and $M$ is the mass parameter\footnote{$M$ is also the conserved charge associated with the Killing vector $\pd_t$ and coincides with the ADM mass if $\lim\limits_{\phi\rightarrow\infty}W(\phi)\text{e}^{Q(\phi)}=1$.} of the black hole as shown in appendix \ref{thermo} to preserve the thermodynamic
relation with the black hole temperature $T$ and entropy $S$,
\[
\quad\quad \quad\quad \quad\quad \quad T=&\b^{-1}=\frac{f'(r_h)}{4\pi}=\frac{\pd_\phi W}{4\pi}\bigg\lvert_{\phi_h} \quad\quad\big(\phi_h\equiv\phi(r_h)\big),\label{temperature}\\
S=&\frac{\phi_h}{4 G_{(2)}},\label{entropy}
\]
where $\rh$ means the location of the outermost horizon.
\subsubsection{Numerical results}
To show the behavior of the QES and its corresponding late-time radiation entropy with respect to $c\cdot G_{(2)}$, we mainly focus on the following concrete cases: (Weyl-related) Witten (or CGHS) black hole \cite{Mandal:1991tz,Witten:1991yr,Callan:1992rs}, (Weyl-related) Schwarzschild black hole\cite{Schwarzschild:1916uq}, (Weyl-related) black hole attractor \cite{Grumiller:2003hq}, JT black hole \cite{Jackiw:1984je,Teitelboim:1983ux}, and AdS-Schwarzschild black hole \cite{Hawking:1982dh,Socolovsky:2017nff}. The first six black holes are (asymptotically) flat\footnote{The curvature for the Weyl-related Witten (CGHS) black hole is zero.} and the prefix "Weyl-related" means that the metric of the theory is related to the original theory by a Weyl transformation (see Appendix \ref{appendix 4: Weyl}). The metrics, dilaton profiles and corresponding $U$, $V$ functions are summarized in Table \ref{tab:black hole models}.
\begin{table}[htbp]
\centering
\caption{Serval eternal black hole solutions in two-dimensional GDT.}
\label{tab:black hole models}
\begin{tabular}{clclclclcl}
\toprule[1pt]
\multicolumn{2}{c}{\multirow{2}{*}{Black hole}}        & \multicolumn{2}{c}{\multirow{2}{*}{$U(\phi)$}}    & \multicolumn{2}{c}{\multirow{2}{*}{$V(\phi)$}}    & \multicolumn{2}{c}{\multirow{2}{*}{$\phi(r)$}}  & \multicolumn{2}{c}{\multirow{2}{*}{$f(r)$}}    \\
\multicolumn{2}{c}{}                           & \multicolumn{2}{c}{}                      & \multicolumn{2}{c}{}                      & \multicolumn{2}{c}{}                      & \multicolumn{2}{c}{}                      \\
\hline
\multicolumn{2}{c}{\multirow{2}{*}{Witten (CGHS)}}      & \multicolumn{2}{c}{\multirow{2}{*}{$\phi^{-1}$}}  & \multicolumn{2}{c}{\multirow{2}{*}{$4\l^2\phi$}}  & \multicolumn{2}{c}{\multirow{2}{*}{$\e^{2\l r}$}}  & \multicolumn{2}{c}{\multirow{2}{*}{$1-\e^{-2\l(r-\rh)}$}}  \\
\multicolumn{2}{c}{}                           & \multicolumn{2}{c}{}                      & \multicolumn{2}{c}{}                      & \multicolumn{2}{c}{}                      & \multicolumn{2}{c}{}                      \\
\multicolumn{2}{c}{\multirow{2}{*}{Weyl-related Witten (CGHS)}}     & \multicolumn{2}{c}{\multirow{2}{*}{$0$}} & \multicolumn{2}{c}{\multirow{2}{*}{$4\l^2$}} & \multicolumn{2}{c}{\multirow{2}{*}{$2\l r$}} & \multicolumn{2}{c}{\multirow{2}{*}{$2\l(r-\rh)$}} \\
\multicolumn{2}{c}{}                           & \multicolumn{2}{c}{}                      & \multicolumn{2}{c}{}                      & \multicolumn{2}{c}{}                      & \multicolumn{2}{c}{}                      \\
\multicolumn{2}{c}{\multirow{2}{*}{Schwarzschild}}   & \multicolumn{2}{c}{\multirow{2}{*}{$\big(2\phi\big)^{-1}$}}   & \multicolumn{2}{c}{\multirow{2}{*}{$2\l^2$}}   & \multicolumn{2}{c}{\multirow{2}{*}{$\l^2r^2$}}   & \multicolumn{2}{c}{\multirow{2}{*}{$1-\frac{\rh}{r}$}}   \\
\multicolumn{2}{c}{}                           & \multicolumn{2}{c}{}                      & \multicolumn{2}{c}{}                      & \multicolumn{2}{c}{}                      & \multicolumn{2}{c}{}                      \\
\multicolumn{2}{c}{\multirow{2}{*}{Weyl-related Schwarzschild}}  & \multicolumn{2}{c}{\multirow{2}{*}{$0$}}  & \multicolumn{2}{c}{\multirow{2}{*}{$2\l^2\phi^{-\frac{1}{2}}$}}  & \multicolumn{2}{c}{\multirow{2}{*}{$2\l r$}}  & \multicolumn{2}{c}{\multirow{2}{*}{$\sqrt{2\l r}-\sqrt{2\l \rh}$}}  \\
\multicolumn{2}{c}{}                           & \multicolumn{2}{c}{}                      & \multicolumn{2}{c}{}                      & \multicolumn{2}{c}{}                      & \multicolumn{2}{c}{}                      \\
\multicolumn{2}{c}{\multirow{2}{*}{Black hole attractor}} & \multicolumn{2}{c}{\multirow{2}{*}{$0$}}   & \multicolumn{2}{c}{\multirow{2}{*}{$4\l^2\phi^{-1}$}}   & \multicolumn{2}{c}{\multirow{2}{*}{$2\l r$}}   & \multicolumn{2}{c}{\multirow{2}{*}{$\log\frac{r}{\rh}$}}   \\
\multicolumn{2}{c}{}                           & \multicolumn{2}{c}{}                      & \multicolumn{2}{c}{}                      & \multicolumn{2}{c}{}                      & \multicolumn{2}{c}{}                      \\
\multicolumn{2}{c}{\multirow{2}{*}{Weyl-related black hole attractor}} & \multicolumn{2}{c}{\multirow{2}{*}{$\phi^{-1}$}}  & \multicolumn{2}{c}{\multirow{2}{*}{$4\l^2$}}  & \multicolumn{2}{c}{\multirow{2}{*}{$\e^{2\l r}$}}  & \multicolumn{2}{c}{\multirow{2}{*}{$2\l\e^{-2\l r}(r-\rh)$}}  \\
\multicolumn{2}{c}{}                           & \multicolumn{2}{c}{}                      & \multicolumn{2}{c}{}                      & \multicolumn{2}{c}{}                      & \multicolumn{2}{c}{}                      \\
\multicolumn{2}{c}{\multirow{2}{*}{JT}}        & \multicolumn{2}{c}{\multirow{2}{*}{$0$}}   & \multicolumn{2}{c}{\multirow{2}{*}{$\frac{2}{L^2}\phi$}}   & \multicolumn{2}{c}{\multirow{2}{*}{$\frac{r}{L}$}}   & \multicolumn{2}{c}{\multirow{2}{*}{$\frac{r^2-\rh^2}{L^2}$}}   \\
\multicolumn{2}{c}{}                           & \multicolumn{2}{c}{}                      & \multicolumn{2}{c}{}                      & \multicolumn{2}{c}{}                      & \multicolumn{2}{c}{}                      \\
\multicolumn{2}{c}{\multirow{2}{*}{AdS-Schwarzschild}}    & \multicolumn{2}{c}{\multirow{2}{*}{$\big(2\phi\big)^{-1}$}}  & \multicolumn{2}{c}{\multirow{2}{*}{$2\l^2+\frac{6}{L^2}\phi$}}  & \multicolumn{2}{c}{\multirow{2}{*}{$\l^2r^2$}}  & \multicolumn{2}{c}{\multirow{2}{*}{$\frac{(r-\rh)(r^2+\rh r+\rh^2+L^2)}{L^2r}$}}  \\
\multicolumn{2}{c}{}                           & \multicolumn{2}{c}{}                      & \multicolumn{2}{c}{}                      & \multicolumn{2}{c}{}                      & \multicolumn{2}{c}{}\\
\bottomrule[1pt]
\end{tabular}
\end{table}

\begin{figure}[htbp]
	\centering
\captionsetup[subfloat]{farskip=10pt,captionskip=1pt}
\subfloat[t][\centering{$Y(r)$ for Witten (CGHS) black hole}]{\label{Fig:Y-CGHS}
			\includegraphics[width =0.45\linewidth]{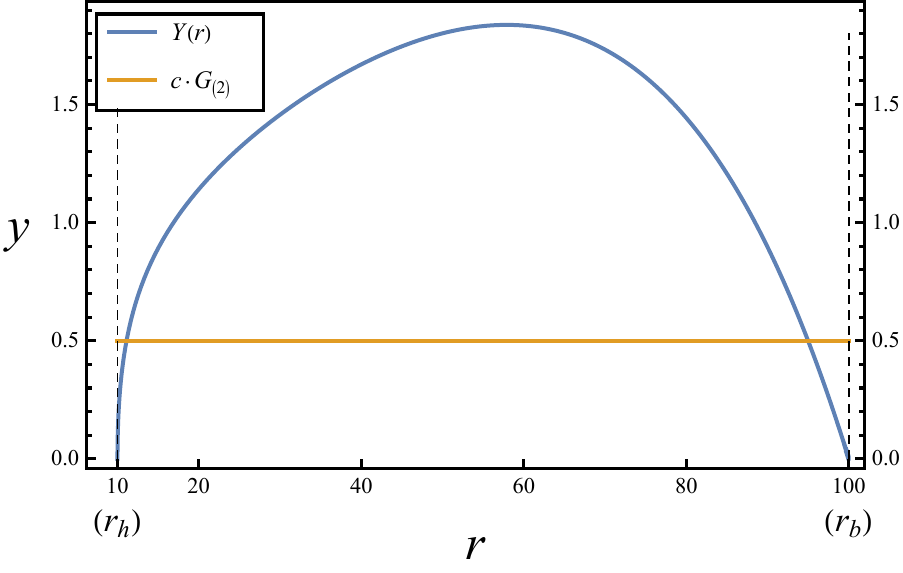}}
\hfill
\subfloat[t][\centering{$Y(r)$ for Weyl-related Witten (CGHS) black hole}]{\label{Fig:Y-WeylCGHS}
			\includegraphics[width =0.45\linewidth]{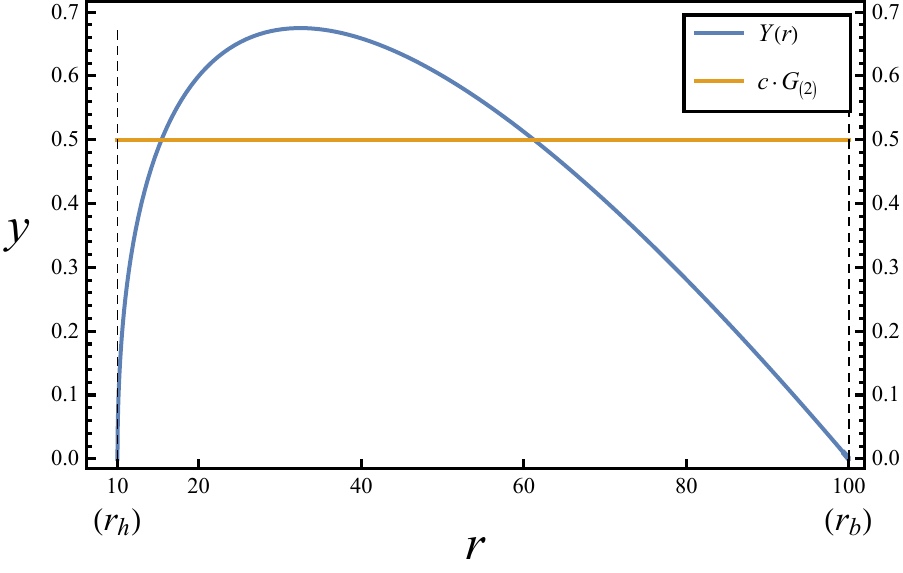}}\\

\subfloat[t][\centering{$Y(r)$ for Schwarzschild black hole}]{\label{Fig:Y-Schwarzshild}
			\includegraphics[width =0.45\linewidth]{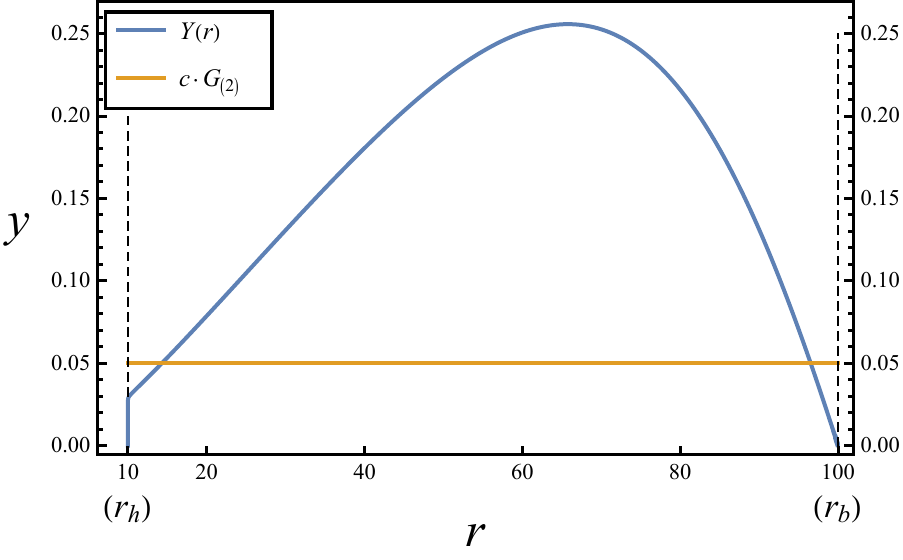}}
\hfill
\subfloat[t][\centering{$Y(r)$ for Weyl-related Schwarzschild black hole}]{\label{Fig:Y-WeylSchwarzshild}
			\includegraphics[width =0.45\linewidth]{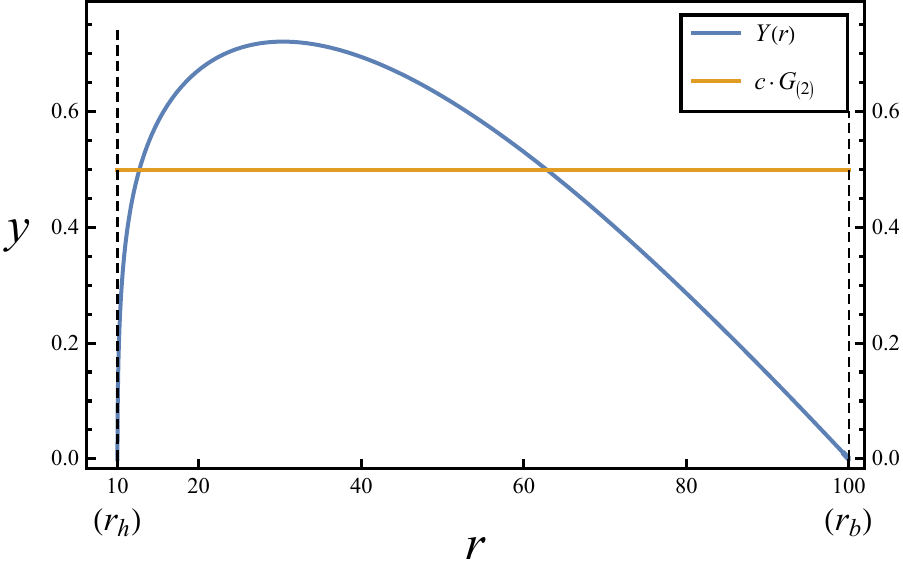}}\\

\subfloat[t][\centering{$Y(r)$ for black hole attractor}]{\label{Fig:Y-BHattractor}
			\includegraphics[width =0.45\linewidth]{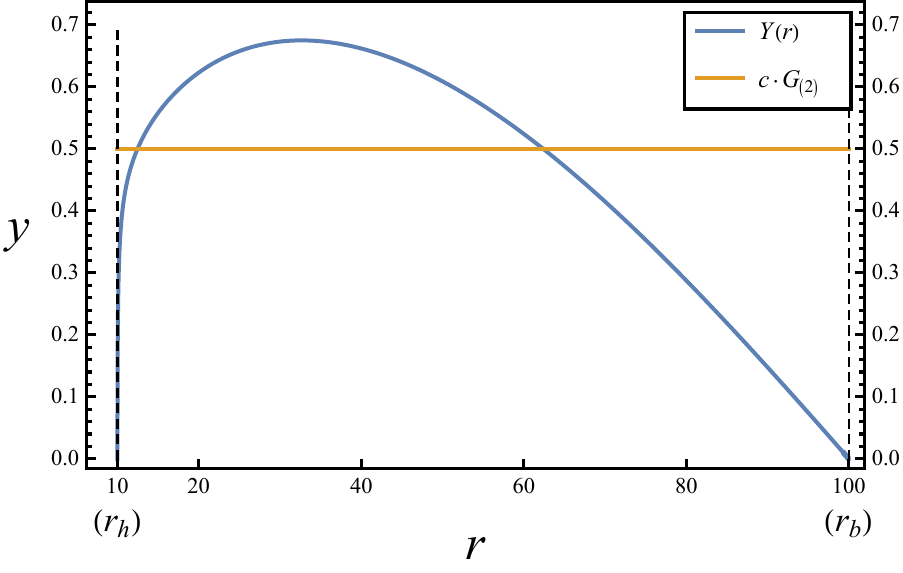}}
\hfill
\subfloat[t][\centering{$Y(r)$ for Weyl-related black hole attractor}]{\label{Fig:Y-WeylBHattractor}
			\includegraphics[width =0.45\linewidth]{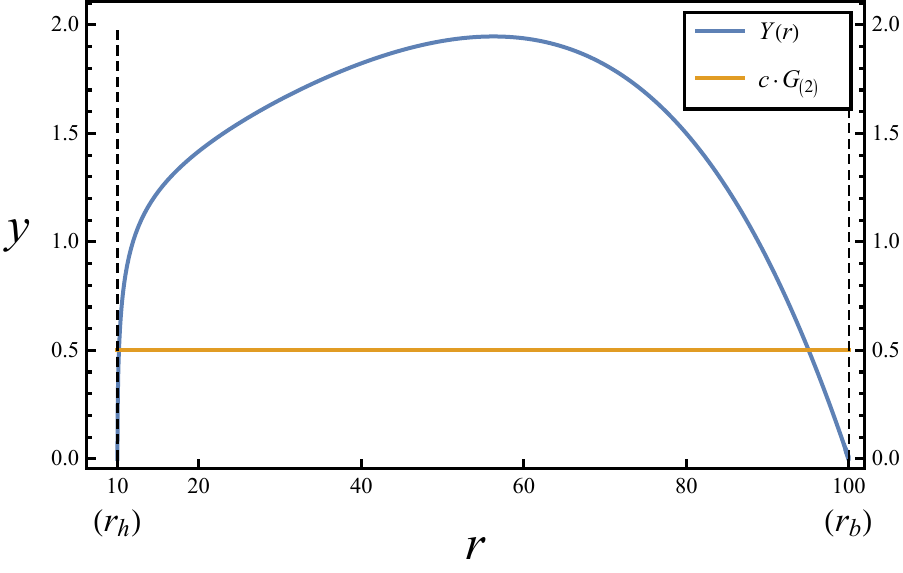}}
	\caption{$Y$-functions \eqref{cG=Y(a)} for (asymptotically) flat black holes. We draw  these diagrams by setting  $\rh=10$, $\l=10^{-2}$, $r_b=10\rh$, }
	\label{Fig: Y function for flat BH}
	\vspace{-0.5em}
\end{figure}

\begin{figure}[htbp]
	\centering
			\includegraphics[width =0.42\linewidth]{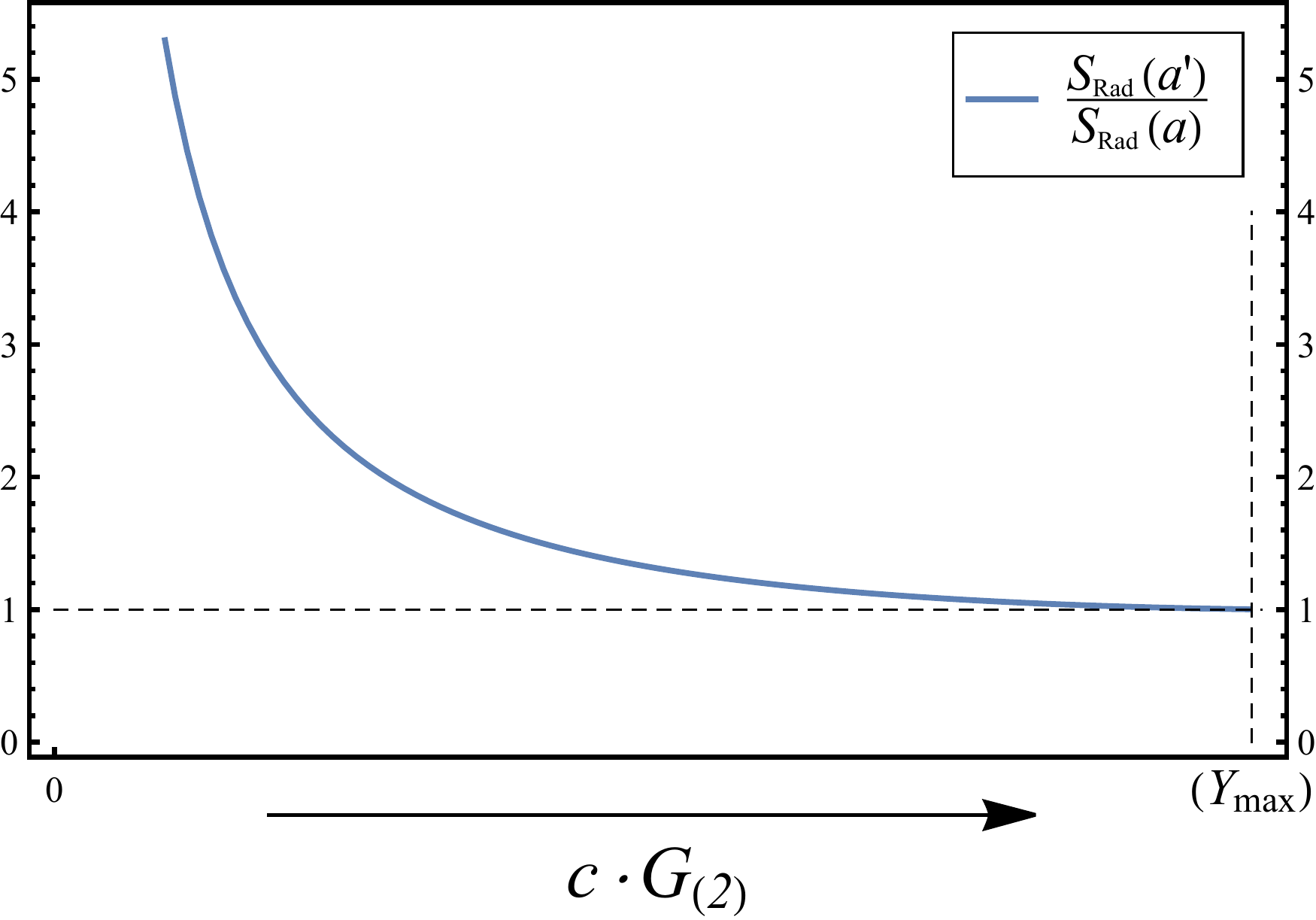}
	\caption{The ratio of the late-time radiation entropy given by the two roots of \eqref{QES equation} for (asymptotically) flat black holes. $a$ represents the root near $\rh$ and $ a'$ represents the root near $r_b$. The numerical result shows that the root near the horizon will always be the boundary of island, if there is one.}\label{Fig:Srad1/Srad2}
	\vspace{-0.5em}
\end{figure}

\begin{figure}[h]
	\centering
\captionsetup[subfloat]{farskip=10pt,captionskip=1pt}
\subfloat[t][\centering{$Y(r)$ for JT black hole}]{\label{Fig:Y-JT}
			\includegraphics[width =0.45\linewidth]{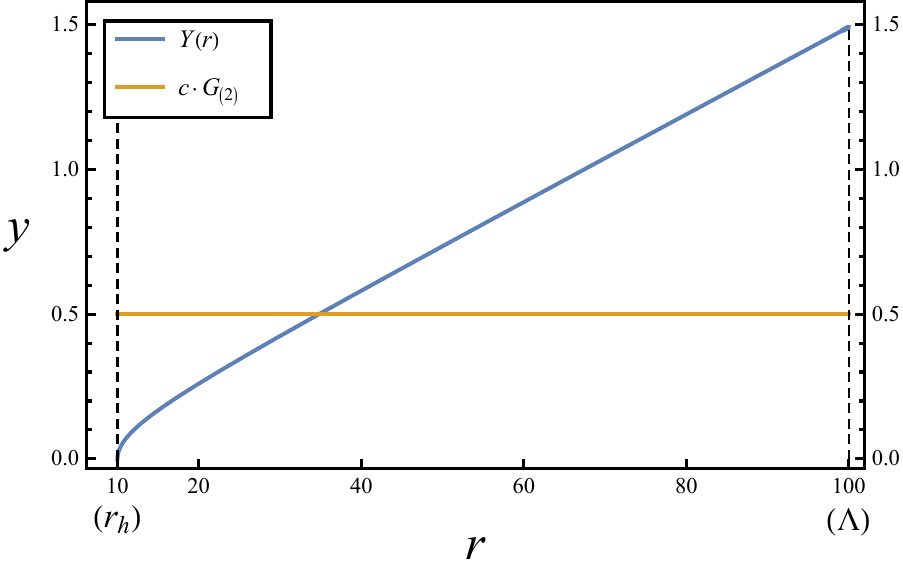}}
\hfill
\subfloat[t][\centering{Enlarged view of Fig.\subref{Fig:Y-JT} at $\rh$}]{\label{Fig:Y-LocalJT}
			\includegraphics[width =0.45\linewidth]{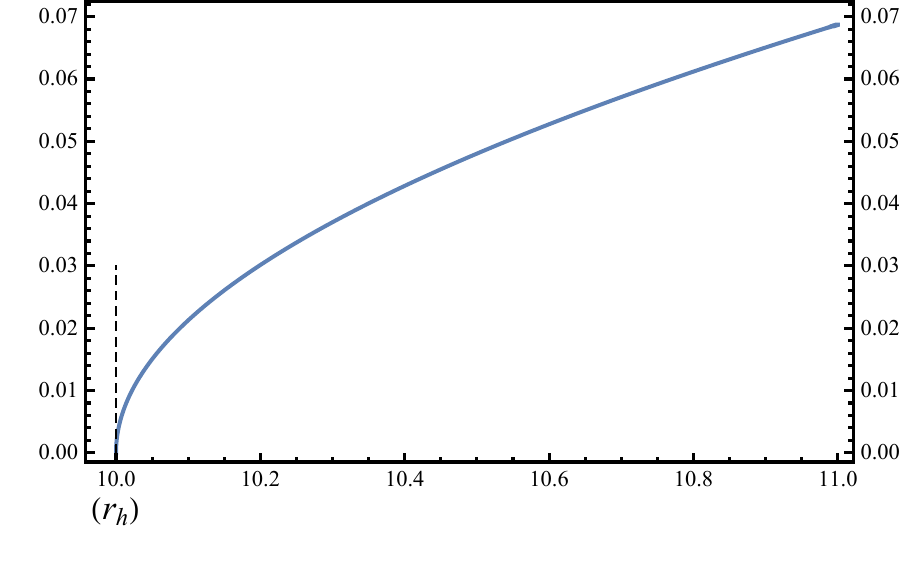}}\\

\subfloat[t][\centering{$Y(r)$ for AdS-Schwarzschild black hole}]{\label{Fig:Y-AdSSchwarz}
			\includegraphics[width =0.45\linewidth]{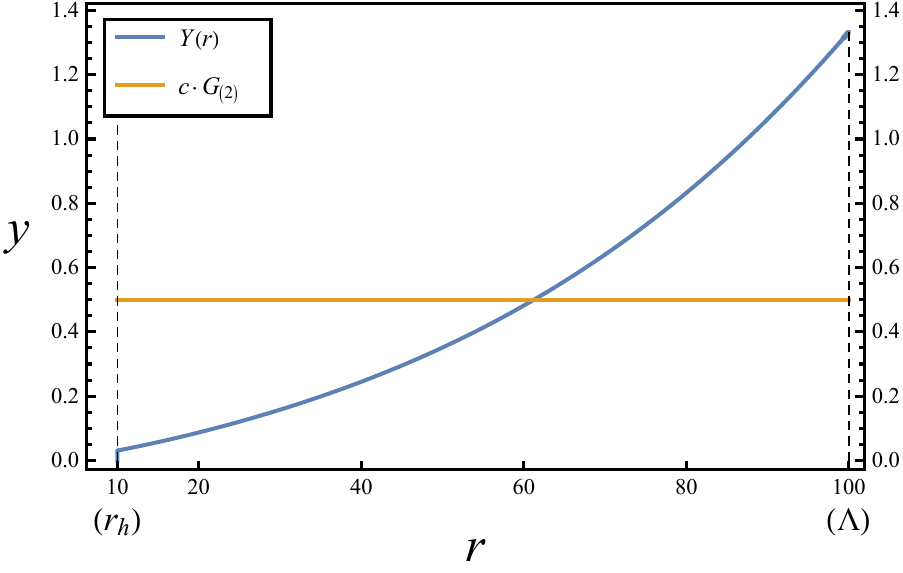}}
\hfill
\subfloat[t][\centering{Enlarged view of Fig.\subref{Fig:Y-AdSSchwarz} at $\rh$}]{\label{Fig:Y-LocalAdSSchwarz}
			\includegraphics[width =0.46\linewidth]{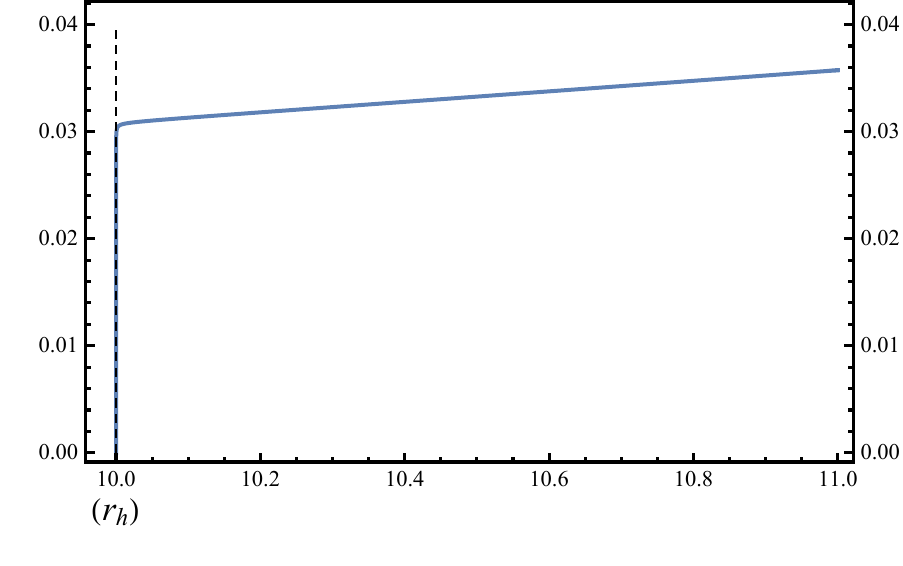}}
	\caption{$Y$-functions \eqref{cG=Y(a)} for (asymptotically) AdS black holes. We draw these diagrams by setting  $\rh=10$, $L=100$, $r^*_b=0$, $\Lambda=10\rh$ ($\l=10^{-2}$ for AdS-Schwarzschild). }
	\label{Fig: Y function for AdS BH}
	\vspace{-0.5em}
\end{figure}

We first draw  $Y$-functions \eqref{cG=Y(a)} for black holes mentioned above. As demonstrated in Fig.\ref{Fig: Y function for flat BH}, for (asymptotically) flat black holes, $Y(r)$ is a concave function that is continuous and consistently greater than or equals to $0$ on the closed interval $[r_h,r_b]$ ($0$ is evaluated at two endpoints). This indicates that when $0<c\cdot G_{(2)}<\YMAXFLAT$, there must be two roots, one  is closer to the horizon (denoted as $a$) and the other  is closer to the boundary of the collecting region (denoted as $a'$). By comparing the late-time radiation entropy given by the two roots, as shown in Fig.\ref{Fig:Srad1/Srad2}, $a'$ is discarded due to the larger entropy given. When $c\cdot G_{(2)}=\YMAXFLAT$, $a$ coincides with $a'$. When $c\cdot G_{(2)}>\YMAXFLAT$, Eq.\eqref{QES equation} has no solution and the island structure is thus destroyed, as stated in conclusion \ref{conclusion 2}.
The situation will be changed for (asymptotically) AdS black holes. As shown in Fig.\ref{Fig: Y function for AdS BH}, $Y(r)$ is monotonically increasing from zero on the interval $[r_h,\Lambda]$, which indicates that eq.\eqref{QES equation} has one and only one root if and only if $0<c\cdot G_{(2)}\leq Y(\Lambda)$. Therefore, as stated in conclusion \ref{conclusion 2}, for (asymptotically) AdS black holes, $c\cdot G_{(2)}$ has an upper bound after the truncation is given. For JT and AdS-Schwarzschild black holes this is given by the value of $Y(r)$ at cut-off.
It can be seen that no matter it is asymptotic flatness (Fig.\ref{Fig: Y function for flat BH}) or asymptotic AdS (Fig.\ref{Fig: Y function for AdS BH}), the behavior of $Y$-function near $r_h$ is similar to that of the square root function $\sqrt{\frac{r}{\rh}-1}$, which is consistent with the previous analysis.

\begin{figure}[t]
	\centering
\captionsetup[subfloat]{farskip=10pt,captionskip=1pt}
\subfloat[t][\centering{$r_{\text{QES}}/\rh$ for asymptotically flat black holes}]{\label{rqesrh}
			\includegraphics[width =0.45\linewidth]{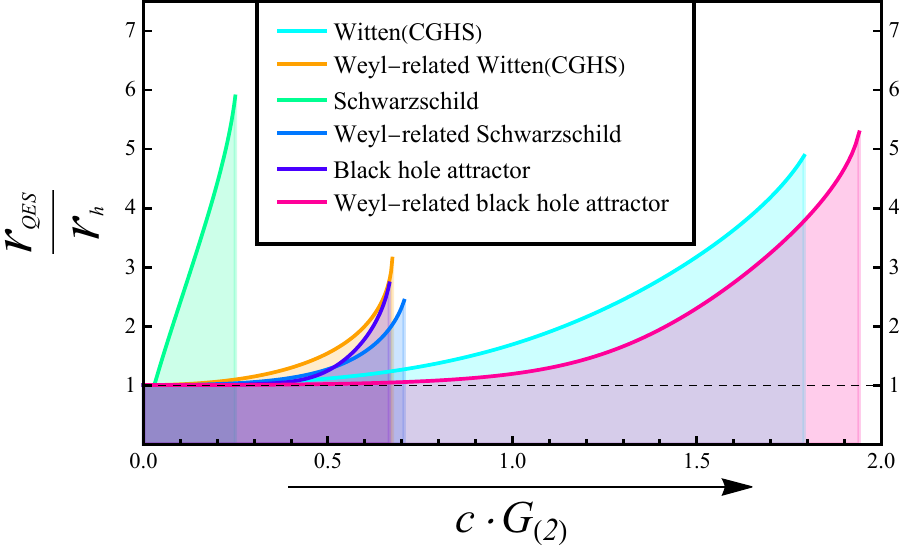}}
~~~~~~~~~~~
\subfloat[t][\centering{Enlarged view of Fig.\subref{rqesrh} at $c\cdot G_{(2)}\sim0$}]{\label{closerqesrh}
			\includegraphics[width =0.44\linewidth]{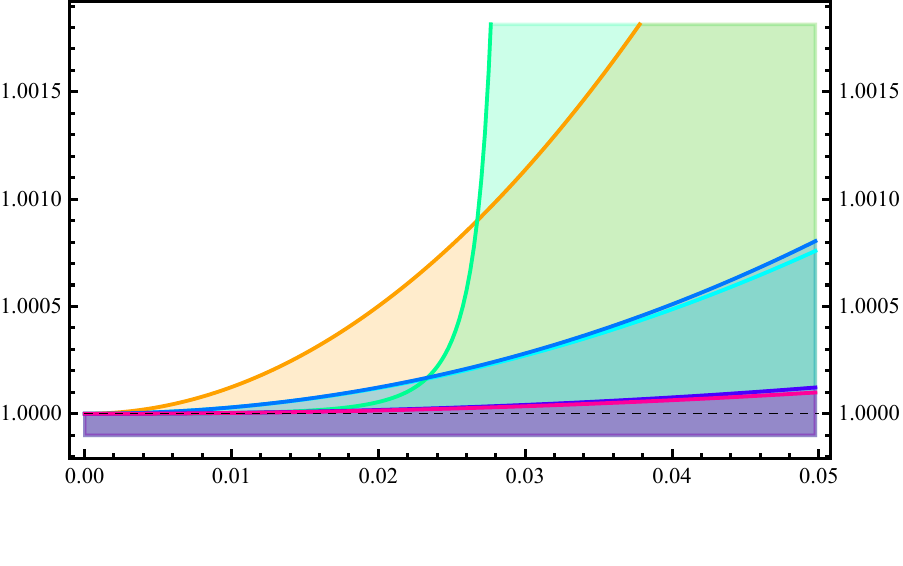}}\\

\subfloat[t][\centering{$S_{\text{Rad}}/2S_{\text{BH}}$ for asymptotically flat black holes}]{\label{SradSBH}
			\includegraphics[width =0.455\linewidth]{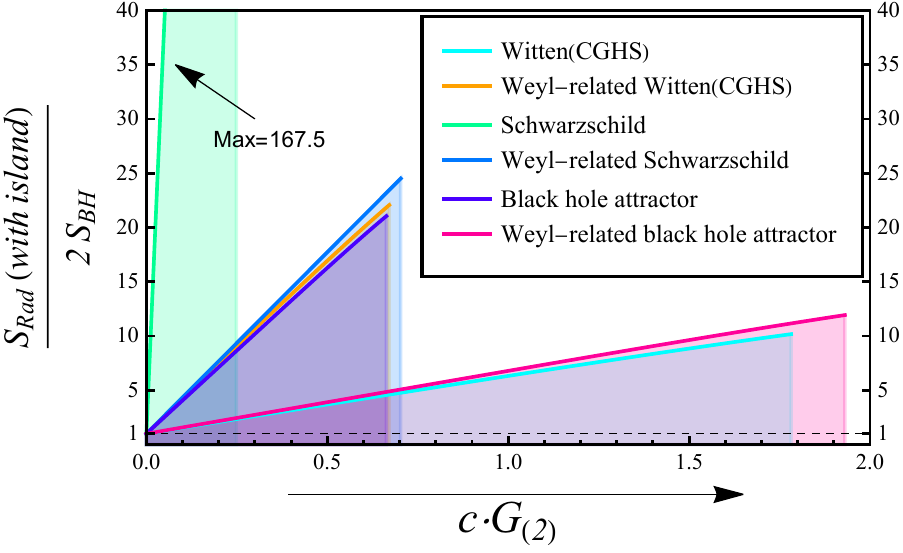}}
~~~~~~~~~~~~
\subfloat[t][\centering{Enlarged view of Fig.\subref{SradSBH} at $c\cdot G_{(2)}\sim0$}]{\label{closeSradSBH}
			\includegraphics[width =0.39\linewidth]{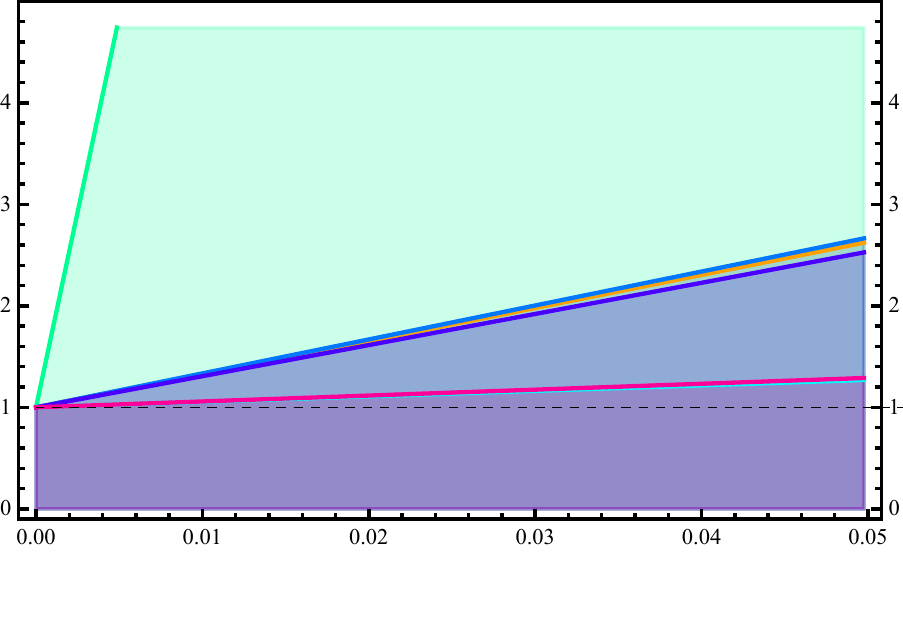}}
	\caption{Curves of the QESs and their corresponding late-time radiation entropy with respect to $c\cdot G_{(2)}$ (for asymptotically flat black holes). Diagrams are plotted with setting  $\rh=10$, $\l=10^{-2}$, $r_b=10\rh$, $G_{(2)}=1/8\pi$.}
	\label{two ratios}
	\vspace{-0.5em}
\end{figure}

\begin{figure}[t]
	\centering
\captionsetup[subfloat]{farskip=10pt,captionskip=1pt}
\subfloat[t][\centering{$r_{\text{QES}}/\rh$ for asymptotically AdS black holes}]{\label{Fig:twoqes}
			\includegraphics[width =0.45\linewidth]{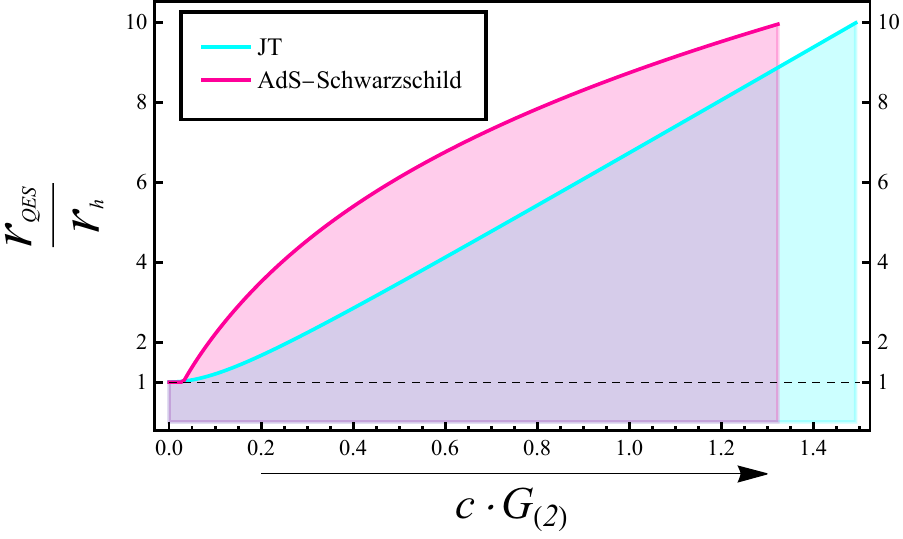}}
~~~~~~
\subfloat[t][\centering{Enlarged view of Fig.\subref{Fig:twoqes} at $c\cdot G_{(2)}\sim0$}]{\label{Fig:closelooktwoqes}
			\includegraphics[width =0.455\linewidth]{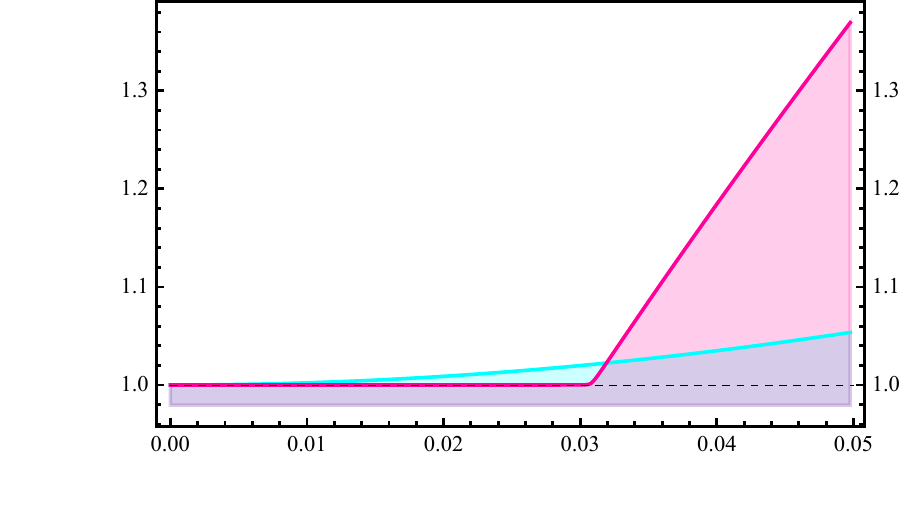}}\\

\subfloat[t][\centering{$S_{\text{Rad}}/2S_{\text{BH}}$ for asymptotically AdS black holes}]{\label{Fig:SredSBHJTAS}
			\includegraphics[width =0.47\linewidth]{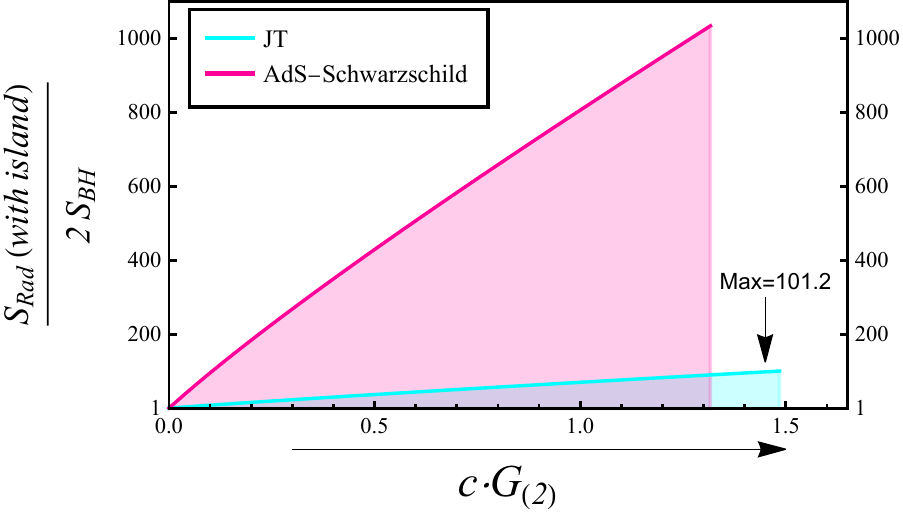}}
~~~~~~~
\subfloat[t][\centering{Enlarged view of Fig.\subref{Fig:SredSBHJTAS} at $c\cdot G_{(2)}\sim0$}]{\label{Fig:LocalSredSBHJTAS}
			\includegraphics[width =0.44\linewidth]{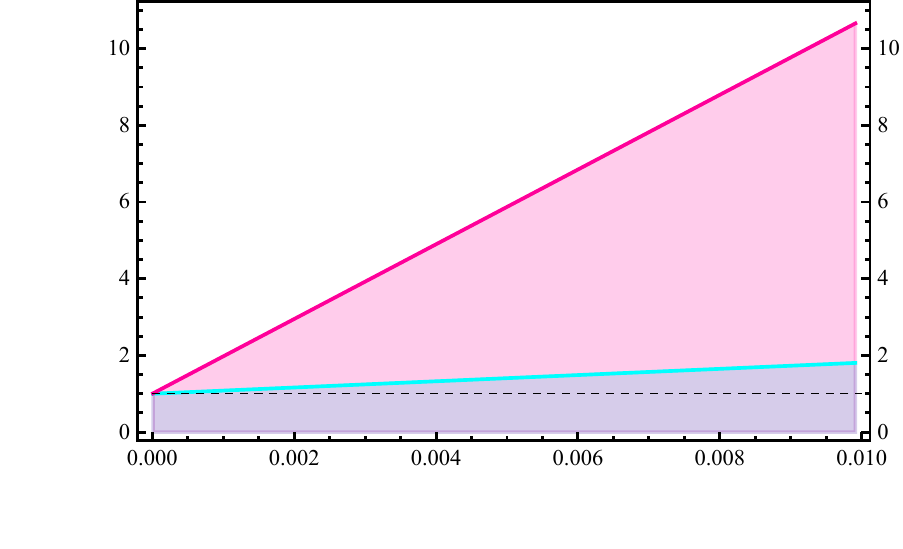}}
	\caption{Curves of the QESs and their corresponding late-time radiation entropy with respect to $c\cdot G_{(2)}$ (for asymptotically AdS black holes). Diagrams are plotted with setting  $\rh=10$, $L=100$, $r^*_b=0$ $\Lambda=10\rh$, $G_{(2)}=1/8\pi$ ($\l=10^{-2}$ for the AdS-Schwarzschild black hole). }
	\label{Fig:ratio for AdS BHs}
	\vspace{-0.5em}
\end{figure}

By solving the intersection of $y= Y(r)$ and $y=c\cdot G_{(2)}$ numerically, we obtain a series of curves of $r_{\text{QES}}$ with respect to $c\cdot G_{(2)}$, see Fig.\subref{rqesrh},(\subref*{closerqesrh}) for (asymptotically) flat black holes and Fig.\subref{Fig:twoqes},(\subref*{Fig:closelooktwoqes}) for (asymptotically) AdS black holes. The corresponding late-time radiation entropy for (asymptotically) flat and AdS black holes are plotted as Fig.\subref{SradSBH},(\subref*{closeSradSBH}) and Fig.\subref{Fig:SredSBHJTAS},(\subref*{Fig:LocalSredSBHJTAS}) respectively.

%

According to the results in Fig.\ref{two ratios} and \ref{Fig:ratio for AdS BHs}, we may summarize the behavior of the quantum extremum surface and its corresponding late-time radiation entropy with respect to $c\cdot G_{(2)}$: When $c\cdot G_{(2)}\ll1$ (or $c\cdot G_{(2)}\sim0$), the location of the QES and the late-time radiation entropy are described by \eqref{approximate solution} and \eqref{late-time entropy: with island} respectively. It results in the QES located in the near-horizon region of the black hole and is a square function of $c\cdot G_{(2)}$, and the radiation entropy is approximately equal to two times the black hole entropy and is a linear function of $c$. When $c\cdot G_{(2)}$ gradually increases, the QES will gradually move away from the  horizon, and the radiation entropy will obviously deviate from the black hole entropy. When $c\cdot G_{(2)}$ grows beyond a certain limit, assuming that $r_b$ (or $\Lambda$ for asymptotically AdS black holes) has been fixed, the equation governing the location of the QES \eqref{QES equation} will have no solution and the island configuration will be destroyed.

\subsection{Island configuration with the backreaction}\label{section: balckreaction}

In the last subsection, we prove that there should be an upper bound on $c\cdot G_{(D)}$ to have an island configuration in the asymptotically flat and $\text{AdS}$ spacetime. When $c\cdot G_{(D)}$ exceeds this upper bound, the island configuration will disappear. This issue has not been discussed in previous literatures \cite{Anegawa:2020ezn,Hartman:2020swn}. However, the calculation in the last subsection is based on the assumption that the $c\cdot G_{(D)}$ is small enough so that the backreaction of the Hawking radiation on the background is negligible. When $c\cdot G_{(D)}$ approaches the upper bound, the effect of Hawking radiation on the background will become significant. In this section, we will reconsider the conditions under which island configuration appears with the finite $c\cdot G_{(2)}$ effect in GDT. The island configuration with the backreaction of the Hawking radiation in JT gravity has been discussed in \cite{Pedraza:2021cvx}.

Consider the two-dimensional dilaton gravity coupled with conformal matter, the Einstein equations can be rewritten as the following formula
\begin{align}
	T_{ab}^\phi+\langle\Psi|T^\chi_{ab}|\Psi\rangle=0,
\end{align}
where $T_{ab}^\phi$ is the stress-energy tensor of dilaton $\phi$ and $T^\chi_{ab}$ is the stress-energy tensor of the conformal matter field. The expression of the second term depends not only on the background metric but also on the quantum state $|\Psi\rangle$. The general form of $\langle\Psi|T^\chi_{ab}|\Psi\rangle$  can be written as
\begin{align}
	&\langle\Psi|T^\chi_{\pm\pm}|\Psi\rangle
	=-\frac{c}{12\pi}\left[(\pd_\pm\rho)^2-\pd_\pm^2\rho\right]
	+\langle\Psi|:T^\chi_{\pm\pm}:|\Psi\rangle,\nn\\
	&\langle\Psi|T^\chi_{\pm\mp}|\Psi\rangle
	=\frac{c}{48\pi}\left(4\pd_+\pd_-\rho+\frac{\lambda_P}{L^2}e^{2\rho}\right)
	\label{Tp},
\end{align}
where $:T_{ab}^\chi:$ is the normal order of $T_{ab}^\chi$ and $\rho$ is the Weyl factor of metric in the conformal gauge $ds^2=e^{2\rho}dx^+dx^-$. $\langle\Psi|:T_{ab}^\chi:|\Psi\rangle$ equals to the total stress-energy tensor $T_{ab}^\chi$  in the locally inertial frame and relates its value in the other coordinates by the local conformal map
\begin{align}
	:T^\chi_{\pm\pm}(x^\pm):
	=\left(\frac{dy^\pm}{dx^\pm}\right)^2:T^\chi_{\pm\pm}(y^\pm):
	-\frac{c}{24\pi}\{y^\pm,x^\pm\},
	\label{conformal-transformation}
\end{align}
where $\{y^\pm,x^\pm\}$ is the Schwarzian derivative. It is easy to show that the stress-energy tensor defined in (\ref{Tp}) gives the correct trace anomaly and the tensor transformation law \cite{Fabbri:2005mw}. The stress-energy tensor defined above can be derived directly from the so-called Polyakov effective action \cite{Polyakov:1981rd}
\begin{align}
	S_{\rm Poly}=&-\frac{c}{24\pi}\int_{\mathcal{M}}\dd^2x\sqrt{-g}\left(\chi R+(\nabla\chi)^2-\frac{\l_P}{L^2}\right)-\frac{c}{12\pi}\int_{\pd\mathcal{M}}\dd x\sqrt{-h}\chi K.\label{poly action}
\end{align}
In this work, we study the island configuration in the eternal black hole geometry. As a result, we assume that all the undetermined functions in the equations of motion are the functions of spatial coordinate $x\equiv\frac{x^++x^-}{2}$. In the conformal gauge,  the equations of motion of the generalized dilaton theories with the Polyakov effective action are
\begin{align}
	&e^{2\r}\pd_\f V(\f)-2\r''-2U(\f)\f''-\pd_\f U(\f)(\f')^2=0,\nn\\
	&\frac{1}{4}U(\f)(\f')^2+\frac{1}{2}\r'\f'-\frac{1}{4}\f''
	=\frac{2cG_{(2)}}{3}\left[\frac{1}{4}\r''-\frac{1}{4}(\r')^2-\t_\pm\right],\label{full-EOM}\\
	&\frac{V(\f)}{4}e^{2\r}-\frac{1}{4}\f''
	=-\frac{cG_{(2)}}{6}\left[\frac{\l}{L^2}e^{2\r}-\r''\right].\nn
\end{align}
where prime means the derivative with respect to $x$ and $\tau_\pm$ is the average of the $\langle\Psi|:T^\chi_{\pm\pm}:|\Psi\rangle$ in some state $|\Psi\rangle$. As discussed in \cite{Almheiri:2019yqk}, we choose the state of the matter field as the Hartle-Hawking state which means that the average of the normal order of the stress-energy tensor vanishes in the Kruskal coordinates $\{\hat{u},\hat{v}\}$,
\begin{align}
	\langle HH|:T^\chi_{\hat{u}\hat{u}}:|HH\rangle=\langle HH|:T^\chi_{\hat{v}\hat{v}}:|HH\rangle=0.
\end{align}
Obviously, we can get the expressions of $\langle HH|:T^\chi_{ab}:|HH\rangle$ in other coordinates by the conformal transformations (\ref{conformal-transformation}) easily. By looking at the equations (\ref{full-EOM}), we find that the background metric in JT gravity and Weyl related Witten black hole are unaffected under the backreaction of the matter field. In these cases, we can solve the full backreaction equations (\ref{full-EOM}) analytically. For the other cases, we will solve the equations (\ref{full-EOM}) in the linear order of $c\cdot G_{(2)}$. Note that the equations (\ref{full-EOM}) satisfy the tensor transformation law, so we shall solve it in the Schwarzschild gauge for simplicity. The undetermined function in this gauge are $\phi(r)$ and $f(r)$ where $f(r)=e^{2\rho(r^*(r))}$.

\subsubsection{Numerical results}
\label{QESwithMatter}
\begin{figure}[htbp]
	\centering
	\captionsetup[subfloat]{farskip=10pt,captionskip=1pt}
	\subfloat[t][\centering{$\pd_{r_a}S_{\rm Rad}$ for Witten (CGHS) black hole}]{\label{Fig:ExtremumW}
		\includegraphics[width =0.7\linewidth]{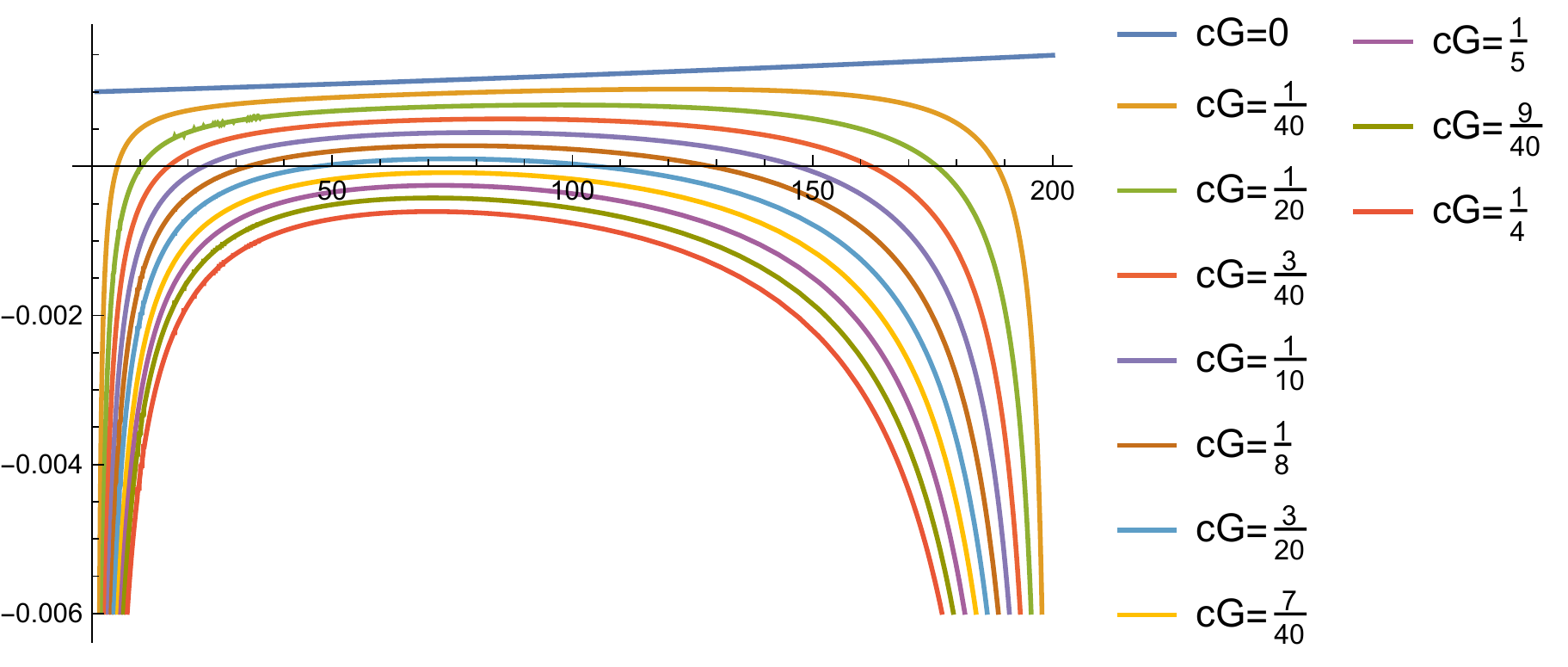}}
	\hfill
	\subfloat[t][\centering{$\pd_{r_a}S_{\rm Rad}$ for Weyl-related Witten (CGHS) black hole}]{\label{Fig:ExtremumWW}
		\includegraphics[width =0.7\linewidth]{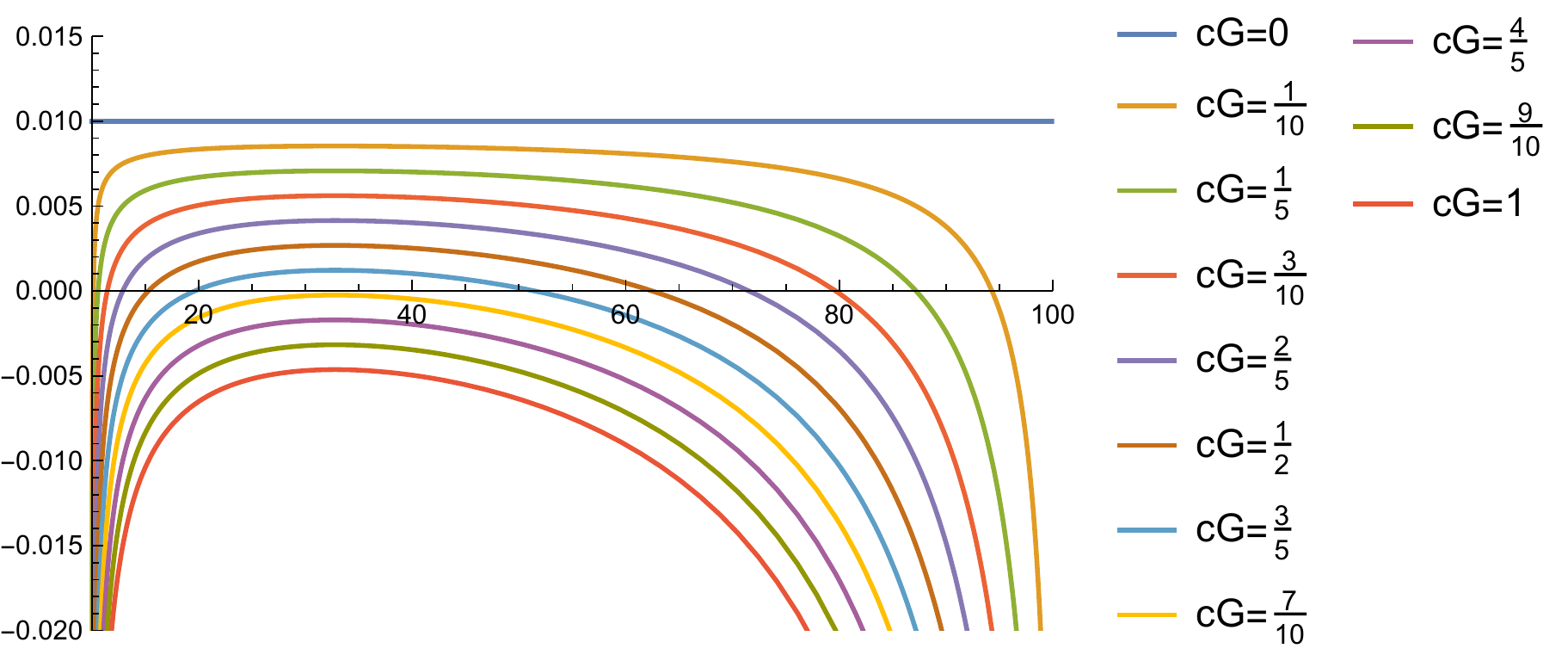}}
	\caption{$r_h=10$, $r_b=200$, $\l_p=L=1$, $\l=10^{-3}$ for Witten black hole and $r_h=10$, $r_b=100$, $\l_p=L=1$, $\lambda=10^{-2}$ for Weyl-related Witten black hole.}
	\vspace{-0.5em}
\end{figure}

For the case of the Weyl-related Witten black hole, the metric is unaffected and the dilaton field can be solved easily. For the case of the Witten black hole, we expand the equations (\ref{full-EOM}) at the linear order of $c\cdot G_{(2)}$ and choose the boundary condition of the perturbative field as
\begin{align}
	&\lim_{r\rightarrow r_h}f_1(r)=0;\quad\quad
	\lim_{r\rightarrow r_h}\pd_rf_1(r)=0;\nn\\
	&\lim_{r\rightarrow\infty}f_1(r)<f_0(r);\quad\quad
	\lim_{r\rightarrow\infty}\phi_1(r)\leq\phi_0(r),
	\label{regular-bdy}
\end{align}
where $f_1(r)$ and $\phi_1(r)$ are the corrections up to the linear order of $c\cdot G_{(2)}$ of the metric function $f(r)$ and dilaton $\phi(r)$, and $f_0(r)$ and $\f_0(r)$ are original solutions without including the backreaction, namely we have
\begin{align}
f(r)=&f_0(r)+c\cdot G_{(2)}f_1(r)+\mathcal{O}\big(c^2\cdot G_{(2)}^2\big),\\
\f(r)=&\f_0(r)+c\cdot G_{(2)}\f_1(r)+\mathcal{O}\big(c^2\cdot G_{(2)}^2\big).
\end{align}
We plot the extreme value conditions (\ref{QES equation}) of  the entanglement entropy in the model of Witten black hole and Weyl-related Witten black hole in Fig.(\ref{Fig:ExtremumW}) and Fig.(\ref{Fig:ExtremumWW}) respectively.
For the Weyl-related Witten black hole, the solutions of $f_1(r)$ and $\phi_1(r)$ are non-perturbative so that the amplitude of $c\cdot G_{(2)}$ doesn't have to be small. We find that the island configuration disappears when $c\cdot G_{(2)}$ exceeds some upper bound from Fig.(\ref{Fig:ExtremumWW}). For the Witten black hole, the amplitude of $c\cdot G_{(2)}$ must be smaller enough to make sure that the perturbation method is applicable. We show in the Fig.(\ref{Fig:ExtremumW}) that, for the island configuration, there indeed exists an upper bound of $c\cdot G_{(2)}$ which is finite and small enough. Specifically speaking, for the parameter chosen as $r_h=10$, $r_b=200$, $\l_p=L=1$, $\l=10^{-3}$, the upper bound of $c\cdot G_{(2)}$ is slightly more than 0.15.

\begin{figure}[htbp]
	\centering
	\captionsetup[subfloat]{farskip=10pt,captionskip=1pt}
	\subfloat[t][\centering{$\pd_{r_a}S_{\rm Rad}$ for Schwarzschild black hole}]{\label{Fig:ExtremumSch}
		\includegraphics[width =0.45\linewidth]{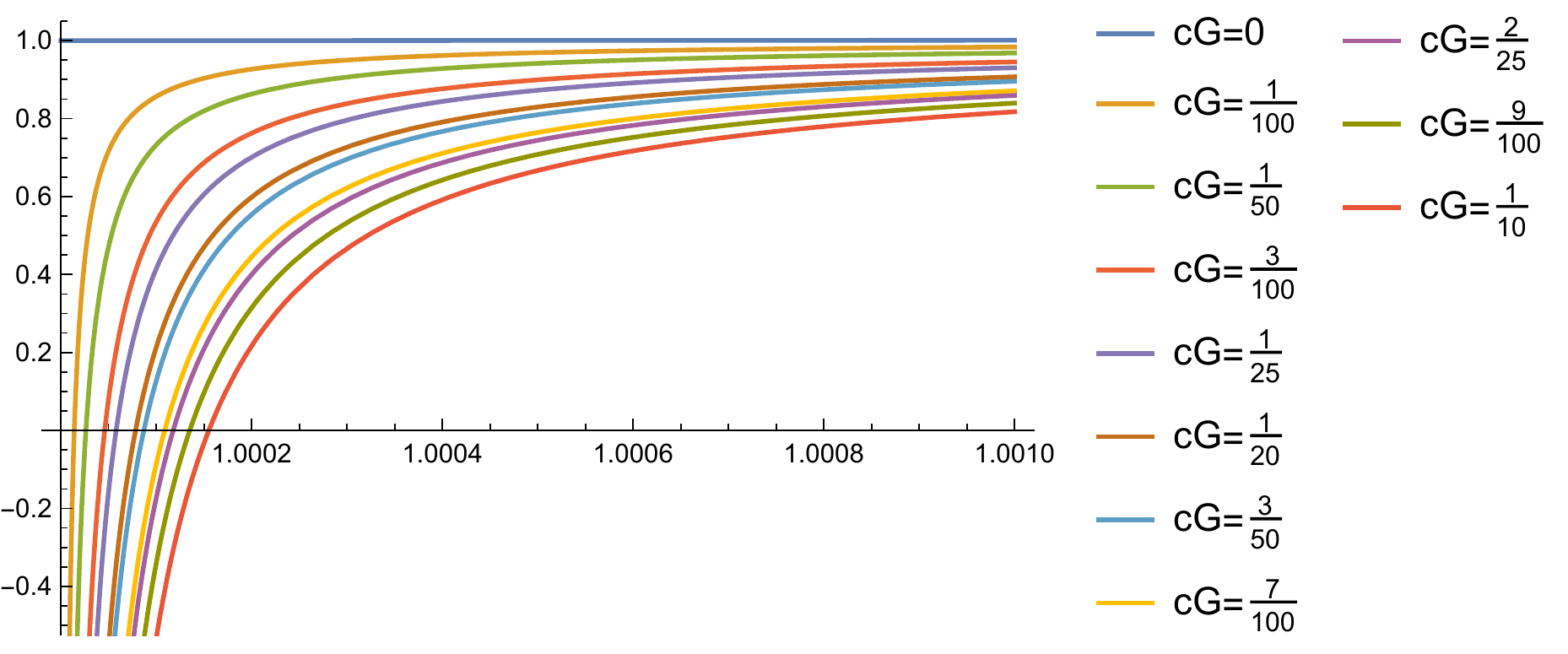}}
	\hfill
	\subfloat[t][\centering{$\pd_{r_a}S_{\rm Rad}$ for Weyl-related Schwarzschild black hole}]{\label{Fig:ExtremumWSch}
		\includegraphics[width =0.45\linewidth]{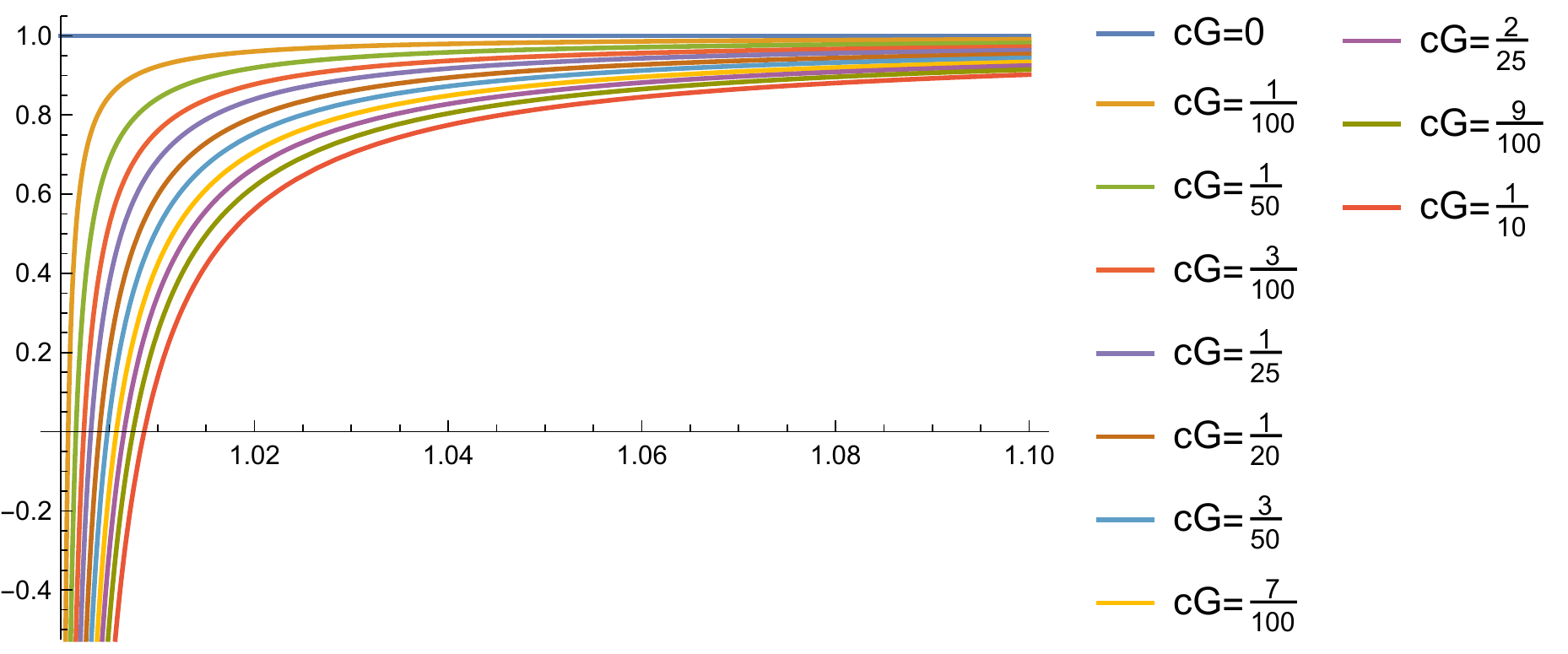}}\\
		\subfloat[t][\centering{$\pd_{r_a}S_{\rm Rad}$ for black hole attractor}]{\label{Fig:ExtremumBHA}
		\includegraphics[width =0.45\linewidth]{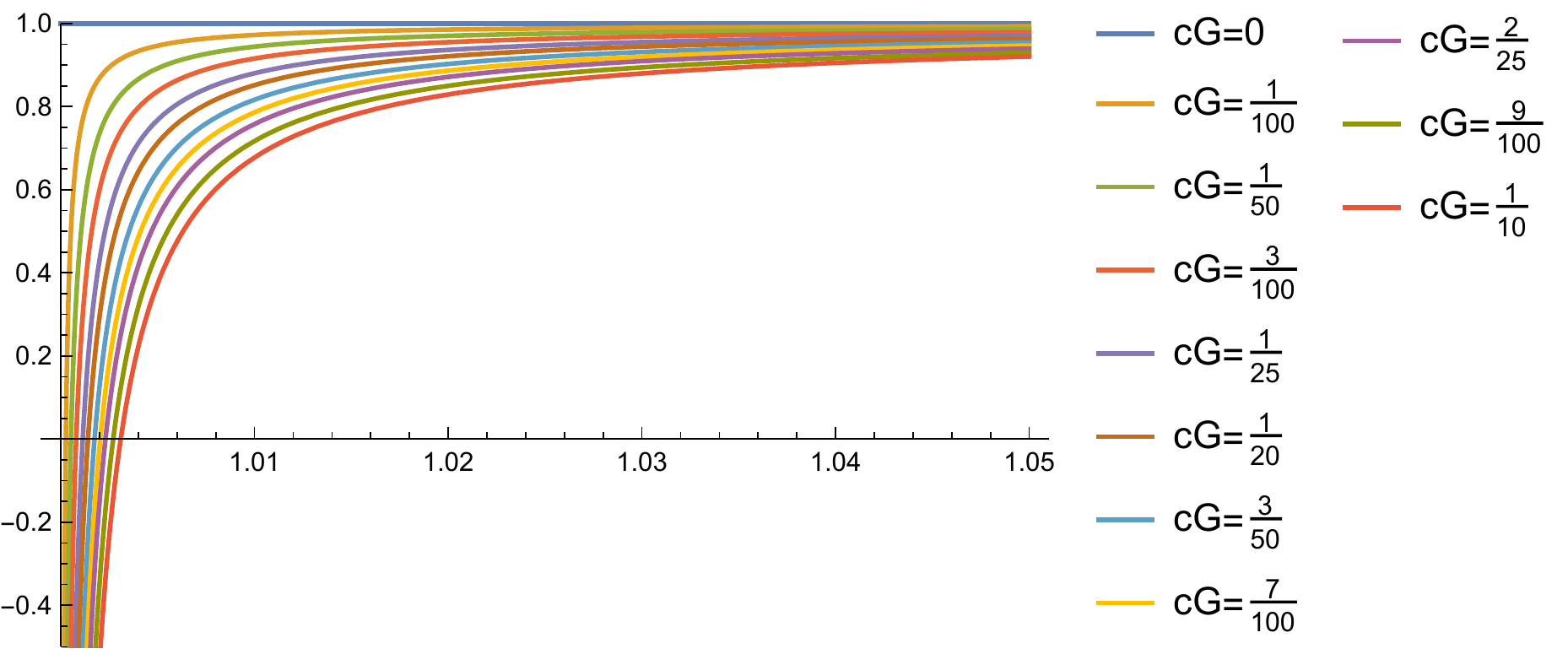}}
	\hfill
	\subfloat[t][\centering{$\pd_{r_a}S_{\rm Rad}$ for Weyl-related black hole attractor}]{\label{Fig:ExtremumWBHA}
		\includegraphics[width =0.45\linewidth]{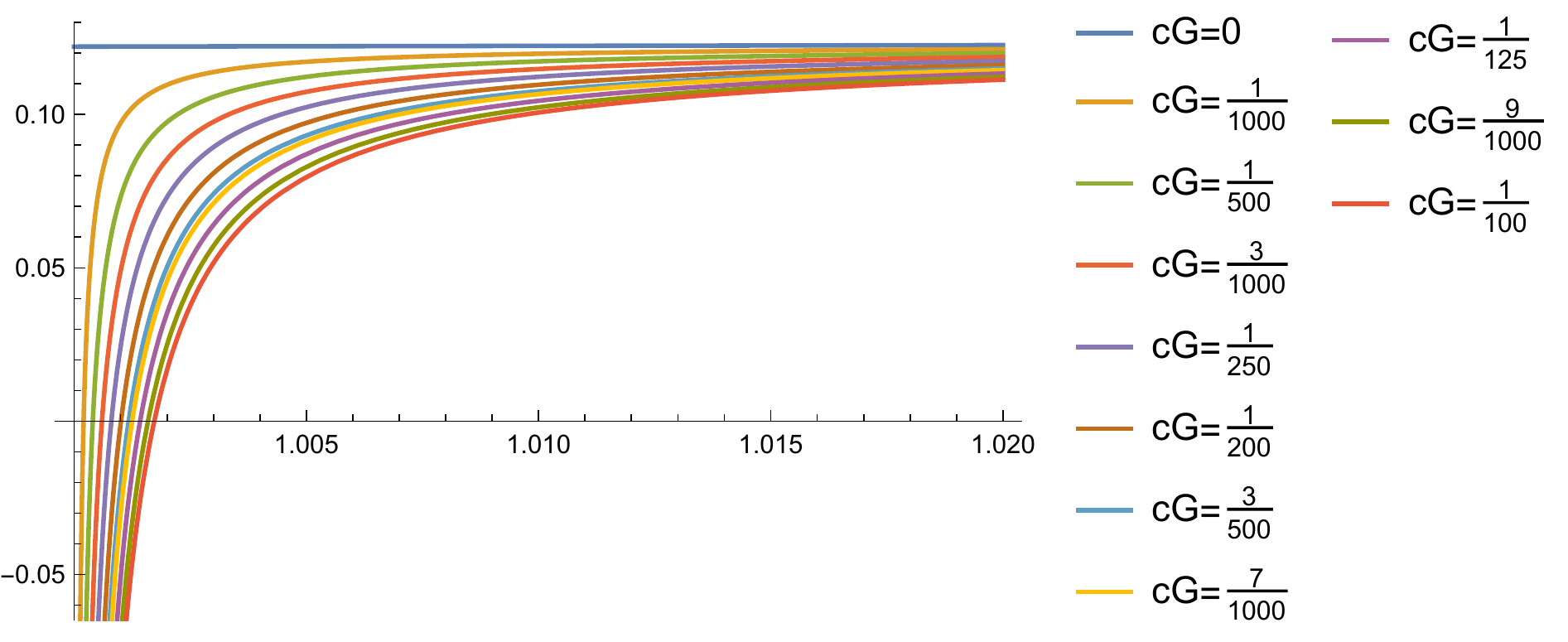}}
	\caption{(a). $r_h=1$, $r_b=10$, $\l_p=\l=L=1$ for Schwarzschild black hole, (b).  $r_h=1$, $r_b=10$, $\l_p=\l=L=1$ for Weyl-related Schwarzschild black hole, (c). $r_h=1$, $r_b=10$, $\l_p=\l=L=1$ for black hole attractor, (d). $r_h=1$, $r_b=10$, $\l_p=L=1$ and $\l=0.1$ for Weyl-related black hole attractor.}
	\vspace{-0.5em}
\end{figure}

For the case of Schwarzschild black hole, Weyl-related Schwarzschild black hole, black hole attractor and Weyl-related black hole attractor, the boundary condition (\ref{regular-bdy}) can not be satisfied, or more specifically, the linear correction terms $f_1(r)$ and $\phi_1(r)$ are dominant and tend to infinity at the asymptotic boundary. However, in order to collect the Hawking radiation, we cut off the bulk geometry at $r_b$ and couple it with flat baths. The linear correction of the metric function $f_1(r)$ and dilaton $\phi_1(r)$ are finite at $r_b$ and can be adjusted small enough when we choose the parameter in the theory appropriately. We display the the root of the equation $\pd_{r_a}S_{\rm Rad}=0$ in Fig.(\ref{Fig:ExtremumSch})-(\ref{Fig:ExtremumWBHA}). It is obvious that there must be a QES near the event horizon which is in agreement with the result asserted in conclusion \ref{conclusion 1}. Unfortunately, We exhaust the parameter space and find that the equation (\ref{QES equation}) always has a root when the $c\cdot G_{(2)}$ is small enough to make sure the perturbation method is applicable.

\begin{figure}[htbp]
	\centering
		\includegraphics[width =0.6\linewidth]{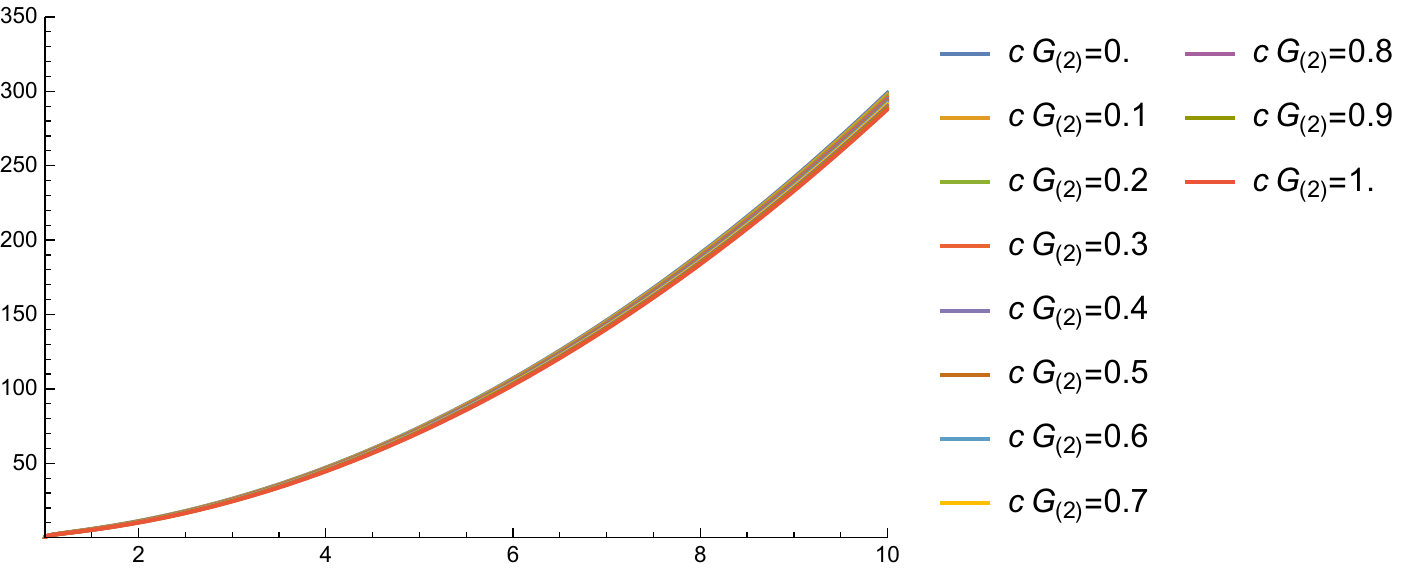}
	\hfill
	\caption{$Y(r)$ for AdS-Schwarzschild black hole with the backreaction. The parameters are chosen as $r_h=\l_p=\l=L=1$, $\Lambda=10r_h$.}
	\vspace{-0.5em}
	 \label{Fig:ExtremumSAdS}
\end{figure}

For the asymptotically AdS black hole, we show that, without the backreaction, the island configuration exists if and only if $0<c\cdot G_{(2)}\leq Y(\Lambda)$ in Fig.(\ref{Fig: Y function for AdS BH}). In this subsection, we reconsider this situation with the backreaction of the matter field. For the JT gravity, the metric and dilaton field is not changed with the matter field when we choose $\lambda_P=1$ which has been discussed in \cite{Pedraza:2021cvx,Almheiri:2014cka}. As a result, the condition of the existence of the island configuration is not changed in JT gravity. For the AdS-Schwarzschild black hole model, we solve the equations (\ref{full-EOM}) at the linear order of $c\cdot G_{(2)}$ with the same boundary conditions (\ref{regular-bdy}). We display our result in the Fig.(\ref{Fig:ExtremumSAdS}).
From the Fig.(\ref{Fig:ExtremumSAdS}), we see that the configuration of the function $Y(r)$ is not sensitive to changes in $c\cdot G_{(2)}$ and the conclusion demonstrated in the Fig.(\ref{Fig: Y function for AdS BH}) remain unchanged.

\section{Conclusions and prospect}\label{sec:3}
In this paper, we systematically study the QES associated with the Hawking radiation {collected by a non-gravitational bath} for general $D$-dimensional ($D\geq2$) asymptotically flat (or AdS) eternal black holes using the island formula. We focus on the non-extremal black hole with spherical symmetry. In this case, the near-horizon geometry is common to all non-extreme black holes and we can use the s-wave approximation in higher dimensional ($D\geq3$) calculating of the matter field entropy. We have obtained the following conclusions:
\begin{itemize}
	\item When $c\cdot G_{(D)}\ll 1$, thanks to the common near horizon structure, there must be a quantum extremal surface (QES) located in the near-horizon region outside the black hole,
	\begin{align}
		r_{\text{QES}}= r_h+ \frac{8\kappa(c\cdot G_{(D)})^2}{9A'(r_h)^2}\exp\Big\{-2\kappa r^*_b-2\rho(r_h)\Big\}+\mo\big((c\cdot G_{(D)})^3\big),
		\label{rqes}
	\end{align}
and the late time radiation entropy saturates $2S_{\text{BH}}$. The formula (\ref{rqes}) is compatible with various known results in \cite{Anegawa:2020ezn,Hashimoto:2020cas,Yu:2021cgi,Nam:2021bml, Wang:2021woy,Kim:2021gzd} and the late time behaviour of the radiation entropy is in good agreement with the previous studies \cite{Almheiri:2019qdq,Penington:2019npb,Almheiri:2019hni,Almheiri:2020cfm,Almheiri:2019psf}.
	\item We go beyond the $c\cdot G_{(D)}\ll1$ limit and scan the parameter space numerically to analyze the location of the QES and its corresponding radiation entropy.
	  \begin{itemize}
		\item[1.] When we ignore the backreaction of the matter field on the background metric, it can be shown generally by the boundedness theorem that there must be an upper bound on $c\cdot G_{(D)}$ to have an island configuration.
		\item[2.] We consider the backreaction of the matter field on the background metric. In the case of Witten black hole and Weyl-related Witten black hole, we can show that for the finite by small enough $c\cdot G_{(D)}$ there is an upper bound on $c\cdot G_{(D)}$ to have an island configuration.
	\end{itemize}
	\item Besides, the numerical results manifest that the location of the QES is just out of the event horizon when $c\cdot G_{(D)}\ll1$. As the value of $c\cdot G_{(D)}$ increases, the QES gradually goes away from the black hole event horizon and the radiation entropy bound will obviously deviate from $2S_{\text{BH}}$.
\end{itemize}

In the section \ref{QESwithMatter}, we discussed the island configuration with the backreaction. Except for the JT gravity and Weyl-related Witten black hole model, we can only get the solution at the linear order of $c\cdot G_{(D)}$. It is important to obtain the full backreaction solution for investigating the condition of the existence of the island configuration. We hope we can resolve this problem numerically in the next step.
It will be interesting to extend our analysis to the near extremal black hole and black hole without spherical symmetry, such as the planar or axisymmetric black hole. Another thing to reconsider is the gravitational effects of the bath in the asymptotically flat black hole because the thermal bath in this case is a gravitational system intrinsically. In asymptotically AdS couple to a gravitating bath, one finds that there is a new saddle point of the bulk geometry in the replica calculation, namely a wormhole connecting the black hole and the gravitational bath\cite{Anderson:2021vof}. After the Page time, this configuration is the dominant contribution and this phenomenon can be regarded as a realization of ER=EPR\cite{Maldacena:2013xja}. {Besides, it is important to proof the QES formula still works in asymptotically flat spacetime with or without some non-gravitational reference system.} The most interesting future problem is to see how an island  is generated dynamically after the page time during the black hole evaporation process. To our knowledge, one can qualitatively reproduce the page curve behavior in several asymptotically flat (or AdS) eternal black holes. However, they can not tell how the black hole information is restored in a concrete way. We are ignorant of the details of the black hole evaporation process even in the semi-classical level. To dynamically generate the island will be an important aspect to reveal the mystery of the black hole information paradox.

\subsection*{Acknowledgements}
We would like to thank Suvrat Raju, Yu-Sen An, Miao He, Shan-Ming Ruan, and Hao Ouyang for the helpful discussion. S.H. would like to appreciate the financial support from Jilin University, Max Planck Partner group as well as Natural Science Foundation of China Grants (No.12075101, No.1204756).
Y.S. is supported by the National Natural Science Foundation of China Grants (No.12105113).

\appendix
\section{Derivation of Eq.\eqref{approximate solution}}\label{appendix 1: derivation of ra}
In this appendix, we present the details  of the derivation of \eqref{approximate solution}. The key is to find the second derivative of the local inverse of $Y$ around $\rh$ (the first derivative is zero), which  can be expressed in terms of the derivative of the primitive function
\[
\left(Y^{-1}\right)''\bigg\lvert_{Y(\rh)}=-\frac{Y''}{\left(Y'\right)^3}\bigg\lvert_{\rh}.
\]
 Firstly, it's useful to set $Y=\frac{3}{2}A' Z^{-1}$, and thereinto,
\[
Z\equiv\a\b+\g,\quad\a\equiv\frac{2\ekr}{f},\quad\b\equiv\frac{\k}{\e^{\k r^*_b}-\ekr},\quad \g\equiv\frac{2\k-f'}{2f}.\label{define of Z}
\]
$Z$ is blow up when $r\rightarrow\rh$, as is evident from the following limits
\[
\lim\limits_{r\rightarrow\rh}\a\sim\lim\limits_{r\rightarrow\rh}\frac{1}{\sqrt{\frac{r}{\rh}-1}}=\infty,\quad\lim\limits_{r\rightarrow\rh}\b=\k\e^{-\k r^*_b}\equiv\b_h,\quad\lim\limits_{r\rightarrow\rh}\g=\frac{-f''(\rh)}{f'(\rh)}\equiv\g_h.
\]
Let's write down the derivative of Y
\[
Y'=&\frac{3}{2}A''Z^{-1}-\frac{3}{2}A'\frac{Z'}{Z^2},\quad Y''=\frac{3}{2}A'''Z^{-1}-3A''\frac{Z'}{Z^2}-\frac{3}{2}A'\bigg(\frac{Z''}{Z^2}-2\frac{(Z')^2}{Z^3}\bigg),
\]
where
\[
Z'=&\a'\b+\a\b'+\g'\nn\\
=&\a f^{-1}\b(\k-f')+\a^2\frac{\b^2}{2}+f^{-1}\Big(-\frac{f''}{2}-\g f'\Big),\label{Z prime}
\]
and
\[
 Z''=&\a''\b+2\a'\b'+\a\b''+\g''\nn\\
 =&\a f^{-2}\b(\k-f')(\k-2f')-\a f^{-1}\b f''+\a^2f^{-1}\b^2(\k-f')\nn\\
 &+\a^2 f^{-1}\b^2\frac{(\k-f')}{2}+\a^3\frac{\b^3}{2}+f^{-2}\Big(f'f''+2\g (f')^2\Big)-f^{-1}\Big(\frac{1}{2}f'''+\g f''\Big).\label{Z prime prime}
\]
In the above we have used the derivatives of $\a$, $\b$, and $\g$ up to the second-order
\[
\a'=&\a f^{-1}(\k-f'),\quad                       &   \a''=&\a f^{-2}(\k-f')(\k-2f')-\a f^{-1}f'',\\
\b'=&\a\frac{\b^2}{2},\quad                       &   \b''=&\a f^{-1}\b^2\frac{(\k-f')}{2}+\a^2\frac{\b^3}{2},\\
\g'=&f^{-1}\Big(-\frac{f''}{2}-\g f'\Big),\quad   &   \g''=&f^{-2}\Big(f'f''+2\g (f')^2\Big)-f^{-1}\Big(\frac{1}{2}f'''+\g f''\Big).
\]
According to (\ref{Z prime}-\ref{Z prime prime}) and the limits of $\a$, $\b$, and $\g$, we have
\[
&~~\frac{Z'}{Z^2}=\frac{\text{e}^{-\k r^*(r)}}{2}\cdot\frac{\b(\k-f')+\mo(\a^{-1})}{\b^2+\mo(\a^{-1})},\\
&~~\frac{Z''}{Z^2} =\frac{\text{e}^{-\k r^*(r)}}{2f}\cdot\frac{\b(\k-f')(\k-2f')+\mo(\a^{-1})}{\b^2+\mo(\a^{-1})},\\
&\frac{(Z')^2}{Z^3}=\frac{\text{e}^{-\k r^*(r)}}{2f}\cdot\frac{\b^2(\k-f')^2+\mo(\a^{-1})}{\b^3+\mo(\a^{-1})},
\]
which leads to the following two limits
\[
&\limrh f\ekr Y''(r)\nn\\
=&-\frac{3}{2}A'(\rh)\limrh\Big(\frac{1}{2}\cdot\frac{\b(\k-f')(\k-2f')+\mo(\a^{-1})}{\b^2+\mo(\a^{-1})}-\frac{\b^2(\k-f')^2+\mo(\a^{-1})}{\b^3+\mo(\a^{-1})}\Big)\nn\\
=&-\frac{3\k^2A'(\rh)}{4\b_h},\\
&\limrh f\ekr\big(Y'\big)^3\nn\\
=&\frac{-3^3}{2^6}\big(A'(\rh)\big)^3\limrh f\text{e}^{-2\k r^*(r)}\cdot\frac{\b^3(\k-f')^3+\mo(\a^{-1})}{\b^6+\mo(\a^{-1})}\nn\\
=&\frac{3^3\k^3}{4^3}\big(A'(\rh)\big)^3\text{e}^{2\rho(\rh)}\b_h^{-3}.
\]
 The second derivative of the local inverse of $Y$ at $\rh$ can thus be obtained by using above limits
\[
\big(Y^{-1}\big)''\Big\lvert_{Y(\rh)}=\limrh\frac{-Y''}{\big(Y'\big)^3}=\k\left(\frac{3}{4}A'(\rh)\exp\big\{\rho(\rh)+\k r^*_b\big\}\right)^{-2}.
\]
In the light of the Taylor expansion of $Y(r)^{-1}$ at $Y(\rh)=0$, we finally arrive at the approximation of $r_a$ upto the second-order of $c^2G_{(D)}^2$
\[
r_a=&\rh+\frac{1}{2}\left(Y^{-1}\right)''\bigg\lvert_{Y(\rh)}\cdot\big(c\cdot G_{(D)}\big)^2+\mo\Big(\big(c\cdot G_{(D)}\big)^3\Big)\nn\\
=&\rh+    \frac{8\k(c\cdot G_{(D)})^2}{9A'(\rh)^2}\exp\Big\{-2\k r^*_b-2\rho(\rh)\Big\}+\mo\Big(\big(c\cdot G_{(D)}\big)^3\Big).
\]
\section{Derivation of Eq.\eqref{late-time entropy: with island}}\label{appendix2:2SBH}
In this appendix, we demonstrate the derivation of \eqref{late-time entropy: with island}. Firstly \eqref{refined radiation entropy} is essentially
\[
S_{\text{Rad}}(r)=\frac{A(r)}{2\GD}+\frac{c}{3}\log d^2(r,r_b)\quad(t=t_b),\label{original radtiation entropy in appendix}
\]
where the first term  is the area term, which is easy to evaluate according to \eqref{approximate solution}
\[
\frac{A(r_a)}{2G_{(D)}}=&\frac{A(\rh)}{2\GD}+\frac{A'(\rh)}{2\GD}(r_a-\rh)+\mo\big((r_a-\rh)^2\big)\nn\\
=&\frac{A(\rh)}{2\GD}+\frac{4\k c^2\cdot G_{(D)}}{9A'(\rh)}\exp\Big\{-2\k r^*_b-2\rho(\rh)\Big\}+\mo(c^3\GD^2).\label{expansion of A}
\]
We then focus on the matter term. Since the approximate behavior of $Y$ near $\rh$ is \eqref{approximation of Y}, meanwhile $Y(r)=-\frac{3}{2}A'(r)\left(\pd_{r}\log d^2(r,r_b)\right)^{-1}$, which gives that the approximate behavior of $\log d^2(r,r_b)$ near $\rh$ is $C_h-\sqrt{\frac{r}{\rh}-1}$, where $C_h\equiv\log d^2(\rh,r_b)$. We can thus obtain its approximation by Taylor expansion of its local inverse function at $\rh$
\[
r=r_h+\frac{1}{2}\Big(\big(\log d^2(r,r_b)\big)^{-1}\Big)''\bigg\lvert_{\rh}\cdot\Big(\log d^2(r,r_b)-C_h\Big)^2.\label{Talor expansion for inverse of logd}
\]
The key step again becomes finding the second derivative of the inverse function at $\rh$
\[
\Big(\big(\log d^2(r,r_b)\big)^{-1}\Big)''=-\frac{\big(\log d^2(r,r_b)\big)''}{\Big(\big(\log d^2(r,r_b)\big)'\Big)^{3}}=-\frac{Z'}{Z^3},
\]
where  $Z$ is defined as \eqref{define of Z}. In terms of the limits obtained in Appendix \ref{appendix 1: derivation of ra}, it's not difficult to find that
\[
\limrh\Big(\big(\log d^2(r,r_b)\big)^{-1}\Big)''=&\limrh\left(\frac{f\e^{-2\k r^*}}{4}\cdot\frac{\b(f'-\k)+\mo(\a^{-1})}{\b^3+\mo(\a^{-1})}\right)\nn\\
=&\frac{1}{4\k}\exp\big\{2\k r^*_b+2\rho(\rh)\big\}.\label{limit of the second derivative of inverse logd}
\]
We then inversely solve for the approximation of $\log d^2(r_a,r_b)$ based on \eqref{Talor expansion for inverse of logd} and \eqref{limit of the second derivative of inverse logd}
\[
\log d^2(r_a,r_b)\approx&C_h-\sqrt{\frac{2(r_a-\rh)}{\Big(\big(\log d^2(r,r_b)\big)^{-1}\Big)''\bigg\lvert_{\rh}}}\nn\\
=&C_h-\frac{8\k c\cdot \GD}{3A'(\rh)}\exp\big\{-2\k r^*_b-2\rho(\rh)\big\}\label{expansion of logd}.
\]
Combine \eqref{expansion of A} with \eqref{expansion of logd},  the final answer arrives
\[
S_{\text{Rad}}(\text{with island})=&\frac{A(r_a)}{2\GD}+\frac{c}{3}\log d^2(r_a,r_b)\nn\\
\approx&\frac{A(r_h)}{2\GD}+\frac{c}{3}\log d^2(r_h,r_b)-\frac{4\k c^2\cdot G_{(D)}}{9A'(\rh)}\exp\Big\{-2\k r^*_b-2\rho(\rh)\Big\}+\mo(c^3\GD^2).
\]
\section{Black hole thermodynamics in GDT}\label{thermo}
In this appendix, we derive the thermodynamic quantities for the 2d dilaton gravity models with action \eqref{GDT's action}.\footnote{Certain assumptions have been made, 1) $\lim\limits_{r\rightarrow+\infty}\phi=+\infty$.  2)$\lim\limits_{\phi\rightarrow+\infty}W(\phi)=+\infty$. 3) $\text{e}^{Q}\neq0$ for finite $\phi$, in this derivation.} We start with the corresponding Euclidean version
\[
I_{\text{E}}=&-\frac{1}{16\pi G_{(2)} }\int_{\mathcal{M}}\sqrt{g}\left(\phi R+U(\phi)\left(\nabla\phi\right)^2+V(\phi)\right)\dd^2x\nn\\
&-\frac{1}{8\pi G_{(2)}}\int_{\pd\mathcal{M}}\sqrt{h} \phi K\dd x+\frac{1}{8\pi G_{(2)}}\int_{\pd\mathcal{M}}\sqrt{h} \lct\dd x,\label{EuclideanAction}
\]
where $\mathcal{M}$ is  spacetime region outside the black hole and the corresponding  boundary $\pd\mathcal{M}$ is $\{r=r_h\}$$\bigcup$$\{r=r_{\text{reg.}}\}$. Note that $\rreg$ is a regulator and should be removed by taking the limit $\rreg\rightarrow\infty$.

 The boundary counterterm $\lct$, as we will see below, should be of form
\[
\lct=\sqrt{W(\phi)\text{e}^{-Q(\phi)}},\label{counterterm}
\]
where the definitions of $W(\phi)$ and $Q(\phi)$ are \eqref{W} and \eqref{Q},respectively. We show this point by reproducing the correct thermodynamics of the black hole.

We start with evaluating the Euclidean action for the black hole solution (\ref{SolutionPhi}--\ref{SolutionMetric}). The bulk contribution reads
\[
I^{\text{bulk}}_{\text{E}}=&\frac{-1}{16\pi G_{(2)}}\int_{\mathcal{M}}\sqrt{g}\bigg\{\phi R+U(\phi)\big(\nabla\phi\big)^2+V(\phi)\bigg\}\dd^2x\nn\\
=&\frac{-1}{16\pi G_{(2)}}\int_{0}^{\beta}\dd\t\int_{\rh}^{\rreg}\dd r\bigg\{-\phi f''(r)+U(\phi)f(r)\left(\frac{\dd\phi}{\dd r}\right)^2+V(\phi)\bigg\}\nn\\
=&\frac{-\beta}{16\pi G_{(2)}}\int_{\phih}^{\phireg}\dd \phi\bigg\{-\phi \pd^2_\phi W+\phi U\pd_{\phi}W+\phi W\pd_\phi U-16\pi G_{(2)}M\phi\pd_{\phi}U\nn\\
&+U\big(W-16\pi G_{(2)}M\big)+\pd_\phi W\bigg\}\nn\\
=&\frac{-\beta}{16\pi G_{(2)}}\bigg\{-\phi\pd_{\phi}W\bigg\lvert^{\phireg}_{\phih}+2W\bigg\lvert^{\phireg}_{\phih}+\phireg U(\phireg)\big(W(\phireg)-16\pi G_{(2)}M\big)\bigg\}.
\]
Let us next consider the on-shell Gibbons-Hawking-York(GHY) term
\[
I^{\text{GHY}}_{\text{E}}=&-\frac{1}{8\pi G_{(2)}}\int_{\pd\mathcal{M}}\sqrt{h}\phi K\dd x\nn\\
=&-\frac{\beta}{8\pi G_{(2)}}\sqrt{f(\rreg)}\cdot\phireg K(\rreg) \nn\\
=&\frac{-\beta}{16\pi G_{(2)}}\bigg\{\phireg\pd_\phi W(\phireg)-\phireg U(\phireg)\big(W(\phireg)-16\pi G_{(2)}M\big)\bigg\}.
\]
It's clear to see that the on-shell bulk term plus the on-shell GHY term equals
\[
\frac{\beta}{16\pi G_{(2)}}\bigg\{2W(\phih)-2W(\phireg)-\phih \pd_{\phi}W(\phih)\bigg\}\label{NoRegulator}.
\]
The above equation is divergent when $\rreg\rightarrow\infty$ since we have assumed that $\lim\limits_{\phi\rightarrow\infty}W(\phi)=\infty$.

The final contribution in \eqref{EuclideanAction} is the boundary counterterm
\[
I^{\text{c.t.}}_{\text{E}}=&\frac{1}{8\pi G_{(2)}}\int_{\pd\mathcal{M}}\sqrt{h}\lct\dd x\nn\\
=&\frac{\beta}{8\pi G_{(2)}}\sqrt{\left(W(\phireg)-16\pi G_{(2)}M\right)\text{e}^{Q(\phireg)}}\cdot\sqrt{W(\phireg)\text{e}^{-Q(\phireg)}}\nn\\
=&\frac{\beta}{8\pi G_{(2)}}\bigg\{W(\phireg)-8\pi G_{(2)}M+\mathcal{O}\big(W(\phireg)^{-1}\big)\bigg\}.\label{onshellcounterterm}
\]
Summing over above contribution and letting $\phireg\rightarrow\infty$, the total on-shell action reads
\[
I^{\text{total}}_{\text{E}}=\beta M-S,
\]
where we have used the definitions of  black hole temperature \eqref{temperature} and Wald entropy \eqref{entropy}. The free energy of black hole in the canonical ensemble reads
\[
\mathcal{F}=-\frac{1}{\beta}\log\mathcal{Z}\sim-\frac{1}{\beta}\log\text{e}^{-I^{\text{total}}_{\text{E}}}=M-TS.
\]
\section{Models related by Weyl transformation}\label{appendix 4: Weyl}
An intriguing feature of the 2d-dilaton gravity model is that one can eliminate (or recover) the kinetic term of the dilaton in the original theory by applying a Weyl transformation \cite{Banks:1990mk,Louis_Martinez_1994,CAVAGLI__2000}. Let's consider a new metric $\hat{g}_{\m\n}$ related to $g_{\m\n}$ by
\[
g_{\m\n}=\text{e}^{-2\Omega}\hat{g}_{\m\n},\quad\Omega=\frac{1}{2}\int^{\phi}U(\phi')\dd\phi'.\label{Weyl transformation}
\]
The bulk term of \eqref{GDT's action} can be re-expressed as follows in terms of $\hat{g}$
\[
\int_{\mathcal{M}}\sqrt{-g}\big(\phi R+U(\phi)\big(\nabla\phi\big)^2+V(\phi)\big)\dd^2 x=&\int_{\mathcal{M}}\sqrt{-\hat{g}}\big(\phi\hat{R}+\text{e}^{-2\Omega}V(\phi)\big)\dd^2 x\nn\\
+&\int_{\pd\mathcal{M}}\sqrt{-\hat{h}}~\phi U(\phi)\hat{n}^\m\hat{\nabla}_\m\phi\dd x,\label{TransformedAction}
\]
where $\hat{R}$ and $\hat{\nabla}$ are Ricci scalar and covariant derivative corresponding to $\hat{g}_{\m\n}$, $\hat{n}^\m$ is the unit vector normal to $\pd\mathcal{M}$. Ignoring the boundary term of no interest, we arrive a simpler theory with vanished kinetic term of dilaton. Three things are noteworthy about the new theory \eqref{TransformedAction}: 1) It gives a linear dilaton solution, i.e., $\phi_{\text{new}}(r)=\text{e}^{-Q_0}r$. 2) The new metric solution is closely related to the original one. We have $f_{\text{new}}(r)\equiv F_{\text{new}}(\phi_{\text{new}})=\text{e}^{-2\Omega}F\lvert_{\phi=\phi_{\text{new}}}$, where $F(\phi)$ is solution of the original theory. 3) The black hole thermodynamic quantities are invariant under the Weyl transformation.\footnote{The reason comes from the following observation: $W(\phi)$ is invariant under the transformation \eqref{Weyl transformation}.}

\bibliographystyle{JHEP}
\bibliography{bibforIslandInEternalBH}{}
\end{document}